\documentclass[11pt,a4paper]{article}
\usepackage{jheppub,amsmath,amssymb,slashed,url,bm,textgreek,upgreek}
\usepackage{jheppub}  
\usepackage{tikz,lipsum,lmodern}
\usepackage[most]{tcolorbox}
\usepackage{amssymb} 
\usepackage{amsmath}
\usepackage{mathtools}
\usepackage{amsfonts}    
\usepackage{dsfont}
\usepackage{pdfpages}
\usepackage{verbatim}
\usepackage{tensor}
\usepackage{mathrsfs}
\usepackage{textgreek} 
\usepackage[mathscr]{euscript}
\usepackage[normalem]{ulem}
\usepackage{tikz}
\usepackage{makecell}
\usetikzlibrary{3d, arrows.meta, decorations.pathreplacing, decorations.markings,calc,shapes.misc,decorations.pathmorphing,patterns.meta, math}

\newcommand{\beq}{\begin{equation}}
\newcommand{\eeq}{\end{equation}}
 
\newcommand{\bea}{\begin{eqnarray}}
\newcommand{\ea}{\end{eqnarray}}
\newcommand{\barr}{\begin{array}}
\newcommand{\earr}{\end{array}}

\def\ie{\begin{equation}\begin{aligned}}
\def\fe{\end{aligned}\end{equation}}

\def\d{{\rm d}}
\def\i{{\rm i}}

\newcommand\U{\text{U}}

\newcommand\SU{\text{SU}}
\newcommand\SO{\text{SO}}

\title{The phase of the gravitational path integral\\
\vspace{-0.9cm}
}

 \author{Xiaoyi Shi and Gustavo J. Turiaci}
 \affiliation{${}^{}$ Physics Department, University of Washington, Seattle, WA, USA}
 \emailAdd{xiaoys5@uw.edu, turiaci@uw.edu}
\abstract{
The gravitational path integral on $S^2 \times S^2$ can be interpreted either as evaluating a contribution to the norm of the Hartle-Hawking wavefunction conditional on spatial $S^1 \times S^2$ topology, or the pair creation rate of black holes in de Sitter. Both interpretations are distinguished at the quantum level. The former requires the path integral to be real and the latter to be imaginary. We develop a formalism to efficiently compute the phase of the gravitational path integral on Einstein spaces. We apply it to a broad class of spacetimes and in particular $S^2\times S^{D-2}$, finding it to be real and positive. We generalize some of the analysis to cases with charge and rotation.  
}

\begin{document}\maketitle

\thispagestyle{empty} 

\setcounter{page}{1}

\newpage

\section{Introduction}

The Hartle and Hawking proposal  \cite{Hartle:1983ai} uses the gravitational path integral on spaces with a positive cosmological constant to prepare the wavefunction of the universe. It instructs us to sum over all smooth geometries that approach a given expanding spatial future geometry. In the case that the spatial slices are 3-spheres, the geometry is complex, combining an early period of Euclidean evolution glued onto an expanding Lorentzian de Sitter solution. In Euclidean signature, this geometry is a 4-sphere. In this context, a reasonable interpretation of the gravitational path integral over the sphere is that it evaluates the norm of the Hartle-Hawking wavefunction. This identification certainly works at the classical level, as already noticed in \cite{Hartle:1983ai}.

\smallskip

The situation at the quantum level is less clear. For pure gravity with a positive cosmological constant, it was realized in \cite{Polchinski:1988ua} that the gravitational path integral on the sphere has a dimension-dependent phase and an overall sign that was recently discussed in \cite{Maldacena:2024spf}. This is incompatible with the claim that the path integral evaluates a norm on a unitary dS Hilbert space, which is expected to be real and positive \cite{Gibbons:1977mu}. A recent proposal by Maldacena, also in \cite{Maldacena:2024spf}, studies whether, when an observer is added to dS beyond pure gravity, the resulting phase of the path integral on Euclidean dS disappears. 

\smallskip

The motivation for the work in this paper is the following. It is interesting to consider the physics of the Hartle-Hawking wavefunction on other spatial  topologies. The simplest case is $S^1 \times S^2$. Surprisingly, the norm of the wavefunction with spatial topology $S^1 \times S^2$ is infinite at the classical level. Quantum effects in the gravitational path integral, that arise when the size of $S^1$ is much larger than the size of the $S^2$, are crucial to render the norm finite \cite{Maldacena:2019cbz,Turiaci:2025xwi}. The contribution to the norm of the Hartle-Hawking wavefunction conditional on a $S^1\times S^2$ spatial topology matches with the small $G_N \Lambda$ approximation of the path integral on $S^2 \times S^2$. This geometry is the Euclidean section of a black hole in dS in its Nariai limit \cite{Nariai,GINSPARG1983245,Bousso:1996au}. 

\smallskip

Can we extend this match beyond the leading order in $G_N \Lambda$? This problem was emphasized recently in \cite{Turiaci:2025xwi}. Motivated by  \cite{Maldacena:2024spf} we will calculate the phase of the gravitational path integral on $S^2 \times S^2$. In this case, there are two possible interpretations of the $S^2 \times S^2$ path integral. The first we already described, as a correction to the norm of the Hartle-Hawking wavefunction. Relatedly, the path integral on $S^D$ or $S^2 \times S^{D-2}$ can also be interpreted as counting microstates of either de Sitter or the Nariai black hole, respectively, although the connection between this picture and the norm of the Hartle-Hawking state is not entirely clear.

\smallskip 

The second interpretation of the $S^2 \times S^2$ path integral is that it computes the decay rate of de Sitter. This was originally proposed by Ginsparg and Perry \cite{GINSPARG1983245}. Following up on the analysis of the instability of gravity on thermal flat space in \cite{Gross:1982cv}, they found that $S^2\times S^2$ has one unstable mode, naively suggesting the path integral is imaginary. This would be the correct result if the path integral is to be interpreted as capturing dS decay and black hole pair creation \cite{Bousso:1995cc,Bousso:1996au}. Perhaps surprisingly, the phase of the path integral on $S^2 \times S^2$ has not been carefully evaluated. It was guessed in \cite{GINSPARG1983245} that it is imaginary based on the presence of one unstable mode, but we learned from \cite{Polchinski:1988ua} that even in the absence of instabilities, the path integral can be complex.

\smallskip

The goal of this paper is to develop a streamlined formalism to evaluate the phase of the gravitational path integral in pure gravity. We will work in $D$ dimensions and mostly focus on the case with a positive cosmological constant. The metric fluctuations around a given geometry can be decomposed into tensor (physical) components, that we take to be transverse and traceless (TT), a vector, and two scalars. We will show that using some classic results from Riemannian geometry one can give a completely general account of the contribution to the phase from the vector and scalar sector. Importantly, the latter contribution is non-trivial due to the presence of negative modes \cite{Polchinski:1988ua}. Therefore, the nontrivial step in the evaluation of the phase is reduced to finding unstable physical TT modes. For example, we find that the gravitational path integral on $S^2 \times S^{D-2}$ is always real and positive regardless of dimension. This is consistent with its interpretation as a correction to the norm of the Hartle-Hawking state, but incompatible with its interpretation as a rate of black hole pair production! Path integrals over Euclidean configurations are required to be imaginary in order to compute vacuum decay in quantum field theory, as illustrated by the conversation between Sagredo and Salviati in \cite{Coleman:1985rnk}. The same feature is presumably true in quantum gravity as well as field theory. 

\smallskip

The fact that the path integral on $S^2 \times S^{D-2}$ is real and positive seems to favor its state-counting interpretation. In a sense, it is reasonable that it does not require introducing an observer since, in Lorentzian signature, this geometry corresponds to a black hole in dS. We present evidence that the path integrals for charged and rotating black holes are also real and positive. Regardless of these results, it is interesting to study the effect an observer can have in these geometries, and we will make some preliminary remarks on this at the end of the paper. This also raises the following question. What is the real rate of black hole pair creation in dS? The Ginsparg-Perry result was used throughout in the literature  \cite{Bousso:1995cc,Bousso:1996au} but as far as we know, there is no rigorous derivation of this rate due to the puzzles on the phase mentioned earlier. Settling this puzzle might have interesting implications for the black hole creation rate during inflation, as well as determining the lifetime of dS.

\smallskip

The organization of the paper is the following. In \textbf{section \ref{sec:QuadraticAction}} we review the derivation of the Einstein action to quadratic order and put it in a form that will be optimal to evaluate quantum corrections. A convenient choice of gauge decouples the scalar modes. In \textbf{section \ref{sec:strategy}} we give a general derivation of the phase of the gravitational path integral in terms of the number of unstable tensor modes, accounting for the vector and scalar modes in full generality. In \textbf{section \ref{sec:Applications}} we apply this method to several examples, including the Nariai black hole $S^2 \times S^{D-2}$ as well as magnetic and rotating generalizations, among others. We discuss future directions in \textbf{section \ref{sec:Discussions}} and leave technical details in appendices. 

\smallskip

\textit{Note added: While this work was in progress, we found out similar calculations were being pursued in \cite{Ivo:2025yek}. We thank the authors for correspondence.}

\section{Quadratic expansion of the Einstein action}\label{sec:QuadraticAction}
The purpose of this section is to simplify the quadratic approximation to the Einstein action in the presence of a positive cosmological constant to its simplest form. We will work on  arbitrary spacetime dimensions $D$. In our conventions, the action is
\beq
I_{\text{EH}} = - \frac{1}{16 \pi G_N} \int_M \d^Dx\,\sqrt{g}\, (R-2\Lambda),
\eeq
where $\Lambda>0$ is the cosmological constant. We will only consider spacetimes without boundaries in this paper, so a Gibbons-Hawking-York term is not necessary but can be easily incorporated. We use our results in this section later in the paper in order to compute the overall phase of the gravitational path integral and its implications.

\subsection{General dimensions}
Our starting point is a space $M$ with $\text{dim}(M)=D$ and metric $g_{\mu\nu}$ that solves the classical equations
\beq\label{eq:EEOM}
R_{\mu\nu} = \frac{2\Lambda}{D-2} g_{\mu\nu},~~~~R= \frac{2 D \Lambda}{D-2}.
\eeq
Geometries with a Ricci curvature that takes this form are called Einstein spaces. We consider metric fluctuations
\beq
g_{\mu\nu} \to g_{\mu\nu} + h_{\mu\nu},
\eeq
and expand the Einstein action to quadratic order
\beq
I_{\text{EH}}[g+h]=I_{\text{EH}}[g] + \delta^2I_{\text{EH}}[h]+ \ldots,
\eeq
where $I_{\text{EH}}[g]$ is the on-shell classical action, $\delta^2I_{\text{EH}}[h]$ is the term quadratic in $h$, and the dots denote higher-order terms in $h$. Since $g_{\mu\nu}$ is a classical solution, there is no linear term in $h_{\mu\nu}$ and $\delta I_{\text{EH}}=0$. An explicit calculation \cite{Gibbons:1978ji,Christensen:1979iy, Yasuda:1983hk} leads to
\bea
\delta^2I_{\text{EH}} &=&\frac{1}{16 \pi G_N} \int_M \d^D x\, \sqrt{g} \Bigg[ \frac{1}{4} h^{\mu\nu} \left(\Delta_L - \frac{4 \Lambda}{D-2}\right) h_{\mu\nu} - \frac{1}{4} h \left(\Delta_0 -\frac{2\Lambda}{D-2}\right)h,\nonumber\\
&&-\frac{1}{2} (\nabla_{\mu} h^{\mu\nu})^2 + \frac{1}{2} \nabla_{\mu}h^{\mu\nu} \nabla_\nu h \Bigg]
\ea
where $h = g^{\mu\nu} h_{\mu\nu}$ is the trace of the metric perturbation with respect to the background metric. The covariant derivatives are also defined with respect to $g_{\mu\nu}$. The Laplacian $\Delta_0 =- \nabla^2$ acts on scalar functions, while the Lichnerowicz operator
\beq
\Delta_L h_{\mu\nu} = - \Box h_{\mu\nu} + R_{\rho \mu} h^{\rho}_{~\nu} + R_{\rho \nu} h^{\rho}_{~\mu} - 2 R_{\rho\mu\sigma\nu} h^{\rho\sigma},
\eeq
acts on rank-2 tensors. This operator and its spectrum will play a central role in the remainder of the paper.
\smallskip

 $\delta^2I_{\text{EH}}$ is invariant under linearized diffeomorphisms $h_{\mu\nu} = \nabla_\mu \xi_\nu + \nabla_\nu \xi_\mu$. We implement a gauge-fixing procedure and follow the standard Faddeev-Popov method. Our choice of gauge-fixing condition is 
\beq
G_\nu[h] = \nabla_\mu h^{\mu}_{~\nu} - \frac{\beta}{2} \nabla_\nu h.
\eeq
The Faddeev-Popov procedure introduces ghosts fields and modifies the Einstein-Hilbert action
\beq
I_{\text{EH}} \to I = I_{\text{EH}} + I_{\text{GF}}+ I_{\text{ghosts}},
\eeq
where
\bea
I_{\text{GF}} &=&\frac{\gamma}{32 \pi G_N}\int_M \d^D x \, \sqrt{g}\, G_\mu[h]\, G^\mu [h].
\ea
In the rest of the paper, we denote by $\delta^2 I$ the term in $I[g+h]$ that is quadratic in $h$. $\gamma$ and $\beta$ are free parameters characterizing our gauge-fixing condition. Gibbons and Perry \cite{Gibbons:1978ji}, as well as \cite{Yasuda:1983hk,Yasuda:1984py,Volkov:2000ih}, observed that the action simplifies considerably when $\gamma$ and $\beta$ satisfy a relation we will derive below. The ghost contribution $I_{\text{ghosts}}$ is written down and analyzed in Appendix \ref{app:ghosts}. Importantly, it does not affect the overall phase of the gravitational path integral. The reason is that the ghosts are intended to reproduce a Jacobian appearing in the implementation of the Faddeev-Popov procedure, which is always positive \cite{Polchinski:1988ua}.

\smallskip

We evaluate the full quadratic action $\delta^2 I$. It is useful to perform a Hodge decomposition of the metric fluctuation \cite{Gibbons:1978ji,Gross:1982cv, Yasuda:1983hk}. We begin with separating pure diffeomorphisms and the trace
\beq\label{eq:ExpHPXH}
h_{\mu\nu} = \phi_{\mu\nu}+ \nabla_\mu \xi_\nu + \nabla_\nu \xi_\mu - \frac{2}{D} g_{\mu\nu} \nabla \cdot \xi + \frac{1}{D} g_{\mu\nu} h.
\eeq
The origin of each individual term in this sum is the following:
\begin{itemize}
    \item The component $\phi_{\mu\nu}$ is defined to be transverse and traceless (TT)
    \beq
    \nabla^\mu \phi_{\mu\nu} =0,~~~~\phi=g^{\mu\nu} \phi_{\mu\nu} = 0,
    \eeq
    and we will refer to it as the tensor or spin-2 sector. These modes solve our gauge-fixing constraint $G_\mu[\phi_{\rho\sigma}]=0$ for any $\beta$, since they are traceless, making their fluctuations physical. 
    \item The middle three terms are parametrized by a vector field $\xi_\mu$, which can be interpreted as a infinitesimal diffeomorphism, and are longitudinal and traceless. We can further decompose that vector field as
    \beq
    \xi_\mu = \eta_\mu + \nabla_\mu \chi,~~~~\nabla \cdot \eta = 0.
    \eeq
    The first term is closed, the second exact. It is known that for compact and orientable manifolds with positive Ricci curvature the first Betti number vanishes, and therefore we do not need to consider harmonic forms in the Hodge decomposition, see theorem 2.4 in \cite{yano1953curvature}.
    
\item The last term is parameterized by a scalar field $h$ and it is simply the trace of the metric fluctuation.
\end{itemize}

The upshot of this analysis is a decomposition of the metric into a tensor mode, a vector mode and two scalar modes
\beq\label{eq:METRICDECOMPF}
h_{\mu\nu} = \underbrace{\phi_{\mu\nu}}_{\text{tensor mode}} + \underbrace{\nabla_{\mu} \eta_\nu + \nabla_\nu \eta_\mu}_{\text{vector mode}} 
 +\underbrace{ 2\left(\nabla_{\mu} \nabla_\nu - \frac{1}{D}g_{\mu\nu} \Box \right) \chi + \frac{1}{D} g_{\mu\nu} h }_{\text{scalar modes}},
\eeq
where $\nabla_\mu \eta^\mu=0$ and $\phi_{\mu\nu}$ is TT. The advantage of this decomposition is that these three sectors are essentially decoupled. This is particularly simple for the tensor and vector modes, since there is only one of each, but the two scalar modes could potentially be coupled with each other. 

\smallskip

It is instructive to first write down the Einstein action to quadratic order, before incorporating the gauge-fixing term, in these new variables \eqref{eq:METRICDECOMPF}. A straightforward calculation gives
\bea
\delta^2 I&=&\frac{1}{16 \pi G_N} \int_M \d^D x\, \sqrt{g} \Bigg[ \frac{1}{4} \phi^{\mu\nu} \left(\Delta_L - \frac{4 \Lambda}{D-2}\right) \phi_{\mu\nu} \nonumber\\
&&~~~~- \frac{D-2}{4D} (h+ 2\Delta_0 \chi)\left[\frac{D-1}{D} \Delta_0 - \frac{2\Lambda}{D-2}\right] (h + 2\Delta_0 \chi) \Bigg],
\ea
see \cite{Yasuda:1983hk}. We shall make two observations on this result. First, it does not involve the vector mode $\eta_\mu$, since it corresponds to a pure diffeomorphism.  Secondly, it only depends on the combination $h+2\Delta_0\chi$ which is also diffeomorphism invariant. To see this, note that in \eqref{eq:ExpHPXH} we explicitly removed the trace of the pure diffeomorphism mode by hand. Therefore, this form of the action makes its symmetries manifest.

\smallskip

We can now incorporate the gauge-fixing terms. Combining the approach of \cite{Yasuda:1983hk} and \cite{Volkov:2000ih}, we obtain
\bea
\delta^2 I&=&\frac{1}{16\pi G_N}\int_M \d^D x \, \sqrt{g}\,\,\Bigg[ \frac{1}{4} \phi^{\mu\nu} \Delta_2 \phi_{\mu\nu}+\frac{\gamma}{2} \eta_\mu \Delta_1^2 \eta^\mu\nonumber\\
&&-\frac{(D-2)}{4 \gamma D^2} h\left( \gamma \widetilde{\Delta}_0-\frac{\gamma^2(D \beta-2)^2}{2(D-2)} \Delta_0\right) h
+\frac{2}{D^2}\chi \Delta_0 \widetilde{\Delta}_0\left( \gamma \tilde{\Delta}_0-\frac{D-2}{2}\Delta_0 \right) \chi.\nonumber\\
&&+\left(1-\frac{2}{D}- \gamma \frac{D\beta-2}{D}\right) h \Delta_0 \widetilde{\Delta}_0 \chi \Bigg]\label{eq:QAHEW}
\ea
In the first line of this action we introduced the differential operator $\Delta_2$ acting on spin-2 TT tensors, and $\Delta_1$ acting on vectors fields, as follows
\bea
\Delta_2\phi_{\mu\nu} &\equiv& \Delta_L\phi_{\mu\nu} - \frac{4\Lambda}{D-2}\phi_{\mu\nu} = -\nabla^2 \phi_{\mu\nu} -2R_{\mu\rho\nu\sigma}\phi^{\rho\sigma},\label{eq:Delta2eee}\\
\Delta_1\eta_\mu &\equiv& - \nabla^2 \eta_\mu - R_{\mu\nu} \eta^\nu.\label{eq:Delta1eee}
\ea
An operator acting on scalars that will play an important role below is 
\beq
\widetilde{\Delta}_0\equiv (D-1) \Delta_0 - \frac{2D \Lambda}{D-2},
\eeq
where $\Delta_0 = -\nabla^2$. In the RHS of equation \eqref{eq:QAHEW}, we have a term quadratic in $h$, quadratic in $\chi$ and also an explicit coupling between $h$ and $\chi$. At this point the freedom in the gauge parameters $\beta$ and $\gamma$ pays off. We can decouple the two scalar modes by simply choosing the prefactor of the $\mathcal{O}(h\chi)$ term to vanish
\beq
1-\frac{2}{D}- \gamma \frac{D\beta-2}{D}=0,~~~~\Rightarrow~~~~\gamma = \frac{D-2}{D\beta-2}.
\eeq
From now on we will make this choice in the rest of the paper and parametrize the gauge-fixing freedom by a single parameter which we take to be $\gamma$. The problem in the scalar sector then reduces to finding the spectrum of the scalar Laplacian.

\smallskip

Let us make the quadratic action \eqref{eq:QAHEW} even more transparent by introducing an inner product between tensors, vectors, and scalars. Following \cite{Volkov:2000ih} we define\footnote{\label{footnote}One can introduce the ``deWitt parameter'' where the integrand of \eqref{eq:normmetr} is replaced by  $h_{\mu\nu} h^{\mu\nu} \to G^{\mu\nu\rho \sigma}h_{\mu\nu} h_{\rho\sigma}$ where $G^{\mu\nu\rho \sigma}=\frac12(g^{\mu\rho}g^{\nu\sigma} + g^{\mu\sigma}g^{\nu\rho}) + \frac{\alpha}{2} g^{\mu\nu} g^{\rho\sigma}$. The analysis in our paper can be easily generalized for values of $\alpha$ leading to a positive-definite metric. A choice of indefinite metrics is also useful in other scenarios \cite{Liu:2023jvm}.}
\bea
\langle h_{\mu\nu}| h^{\mu\nu} \rangle &=& \frac{1}{32\pi G_N} \int \d^Dx\,\sqrt{g}\, h_{\mu\nu}h^{\mu\nu},\label{eq:normmetr}\\
\langle \xi_{\mu}| \xi^\mu \rangle &=& \frac{1}{32\pi G_N} \int \d^Dx\,\sqrt{g}\, \xi_\mu \xi^\mu,\label{eq:normvect}\\
\langle \chi| \chi \rangle &=& \frac{1}{32\pi G_N} \int \d^Dx\,\sqrt{g}\, \chi^2 \label{eq:normsca}
\ea
This inner product will also appear below in the definition of the ultralocal measure for the metric. We can write compactly the bosonic action as
\bea
\delta^2 I &=& \frac{1}{2} \langle \phi^{\mu\nu} | \Delta_2 \phi_{\mu\nu} \rangle + \gamma \langle \eta_\mu | \Delta_1^2 \eta^\mu\rangle+\frac{4}{D^2} \langle \chi | \Delta_0 \widetilde{\Delta}_0 \widetilde{\Delta}_0^\gamma \chi \rangle - \frac{D-2}{2\gamma D^2 } \langle h | \widetilde{\Delta}_0^\gamma h\rangle. \label{eq:quadact}
\ea
We have introduced our final differential operator
\beq
\widetilde{\Delta}_0^\gamma = \gamma \widetilde{\Delta}_0 -\frac{D-2}{2}\Delta_0,
\eeq
which depends on the choice of gauge fixing. We presented the final expression for $\delta^2 I$ in a similar spirit as the formalism of \cite{Volkov:2000ih}, but extended to arbitrary dimensions.

\smallskip

We can observe from \eqref{eq:quadact} that the path integral in the short-distance modes of $\chi$ will converge, while those in the $h$ integral are all divergent. This is the general phenomenon observed by Gibbons, Hawking, and Perry in \cite{Gibbons:1978ac}. At short distances, much smaller than the scale set by the cosmological constant, all the differential operators that appear in the scalar action (as well as tensor and vector sectors) are positive. This holds since $\widetilde{\Delta}_0^\gamma$ and $\widetilde{\Delta}_0$ are approximated in this regime by
\beq
\widetilde{\Delta}_0 \approx (D-1) \Delta_0,~~~~\widetilde{\Delta}_0^\gamma \approx \frac{2\gamma(D-1) -(D-2)}{2} \Delta_0.
\eeq
Since $\Delta_0$ is positive for non-trivial modes, this guarantees that $\widetilde{\Delta}_0$ and $\widetilde{\Delta}_0^\gamma$ are both positive at short distances as long as $\gamma$ is not too small. Indeed we will only consider the range
\beq\label{eq:gammarange}
\gamma>\frac{D-2}{2(D-1)}.
\eeq
Therefore, the $\chi$ action is strictly positive, while the $h$ action is strictly negative in this regime. If we consider compact spacetimes, this conclusion actually holds for all but a finite number of long-wavelength modes. 

\smallskip

In the range \eqref{eq:gammarange} for $\gamma$, the parameter $\beta$ covers the interval
\ie
\frac{2}{D}<\beta <2
\fe
where the upper bound is achieved at $\gamma=(D-2)/2(D-1)$ and the lower bound is achieved as $\gamma\to \infty$. As long as we restrict ourselves to $\gamma$ and $\beta$ within this range and generic, the calculation of the phase and the overall one-loop determinant will be the most straightforward. This is further explained in Appendix \ref{app:ghosts}.

\subsection{The path integral measure}\label{sec:PIM}

Since we replaced the path integral over $h_{\mu\nu}$ by a path integral over $\phi_{\mu\nu}$, $\eta_\mu$, $\chi$ and $h$, it is important to determine the integration measure over these new variables. In particular, not all modes in $\eta_\mu$ and $\chi$ will necessarily appear in the integral path, since some of them do not generate any nontrivial $h_{\mu\nu}$. 

\smallskip

The integration measure over $h_{\mu\nu}$ is taken to be the ultralocal one, implicitly defined through
\beq
\int Dh\, e^{- \mu_0^2\langle h_{\mu\nu}| h^{\mu\nu}\rangle}=1,
\eeq
where the inner product appearing in the exponential was defined in equation \eqref{eq:normmetr}. $\mu_0$ is a dimensionfull parameter that will not be important in our discussion. Let us now decompose our metric as in \eqref{eq:METRICDECOMPF} and replace it into the inner product
\beq\label{eq:measure}
\langle h_{\mu\nu} | h^{\mu\nu} \rangle = \langle \phi_{\mu\nu} | \phi^{\mu\nu} \rangle + 2 \langle \eta_\mu | \Delta_1 \eta^\mu \rangle + \frac{4}{D}\langle \chi | \Delta_0 \tilde{\Delta}_0 \chi\rangle + \frac{1}{D} \langle h | h\rangle.
\eeq
This formula can be derived in a straightforward fashion by replacing \eqref{eq:METRICDECOMPF} into the definition of the metric norm \eqref{eq:normmetr} and integrating by parts, using the Ricci formula for the commutator of covariant derivatives. We can now extract the integration measure over $\phi_{\mu\nu}$, $\eta_\mu$, $\chi$ and $h$ by replacing the decomposition of the inner product in the definition of the ultralocal measure
\beq\label{eq:measuredef}
\int D \phi_{\mu\nu} \, e^{-\mu_0^2\langle \phi_{\mu\nu} | \phi_{\mu\nu} \rangle } \int D \eta_\mu \, e^{-2\mu_0^2\langle \eta_\mu | \Delta_1 \eta^\mu\rangle} \int D\chi \, e^{-\frac{4\mu_0^2}{D}\langle \chi | \Delta_0 \widetilde{\Delta}_0 \chi\rangle} \int D h\, e^{- \frac{\mu_0^2}{D}\langle h| h\rangle} = 1.
\eeq
The measures extracted from this formula are trivial for $\phi_{\mu\nu}$ and $h$ but contain nontrivial functional determinants in the $\eta_\mu$ and $\chi$ sectors. They are the only modes in \eqref{eq:METRICDECOMPF} that appear with derivatives acting on them. 

\smallskip

\paragraph{Comment on vector modes} Not all modes over $\eta_\mu$ need to be integrated. Since the associated metric fluctuation is $h_{\mu\nu} \supset \nabla_{\mu}\eta_{\nu}+\nabla_{\nu}\eta_{\mu}$ this means that modes satisfying 
\beq\label{eq:KVE}
\nabla_\mu \eta_\nu + \nabla_\nu \eta_\mu =0,
\eeq
should be excluded. These are precisely Killing vectors of the underlying spacetime, and notice that \eqref{eq:KVE} automatically forces $\nabla_\mu \eta^\mu =0$. Since the original integral is over $h_{\mu\nu}$, these $\eta_\mu$ modes should be excluded by hand. Also, since the inner product becomes, after integrating by parts, proportional to $ \langle \eta_\mu | \Delta_1 \eta^\mu\rangle $, this implies that Killing vectors satisfy
\beq
\Delta_1 \eta_\mu =0,
\eeq
and are therefore in the kernel of $\Delta_1$. This also implies that $\Delta_1$ is a non-negative operator, we will come back to this in the next section.

\smallskip

\paragraph{Comment on scalar modes} Let us consider the scalar mode $\chi$. Given their associated metric fluctuations, profiles that satisfy
\beq
\left(\nabla_\mu \nabla_\nu - \frac{1}{D} g_{\mu\nu} \Box\right) \chi =0,
\eeq
should be excluded. This also has a nice geometric interpretation which should be clear from the way this scalar mode was originated. Construct a vector field as
\beq\label{eq:SCKV}
\xi_\mu = \nabla_\mu \chi.
\eeq
The excluded modes satisfy
\beq\label{eq:CKVSS}
\nabla_\mu \xi_\nu + \nabla_\nu \xi_\mu = \frac{2}{D} g_{\mu\nu} \nabla_\rho \xi^\rho.
\eeq
This implies that $\xi_\mu$ is a  conformal Killing vector (CKV), since the right hand side of \eqref{eq:CKVSS} is non-zero and proportional to the metric tensor.\footnote{All CKV that are not themselves Killing vectors are pure gradients. This is because, taking \eqref{eq:CKVSS} as a starting point, we can apply the Hodge decomposition to $\xi_\mu$ (which for Einstein spaces with positive $\Lambda$ does not include harmonic forms). The divergence-free contribution to the solution is necessarily a Killing vector and therefore the only possible source of non-isometric CKV is from the longitudinal mode.} Since the contribution to the metric inner product is proportional to $ \langle \chi| \Delta_0 \widetilde{\Delta}_0 \chi \rangle$, these modes also are annihilated by $\Delta_0 \widetilde{\Delta}_0$. Since the only zero-mode of $\Delta_0$ are constant functions, the CKV satisfy
\beq
\widetilde{\Delta}_0 \chi = 0.
\eeq
 This argument also implies  $\Delta_0 \widetilde{\Delta}_0$ is non-negative, we will come back to this later.

\smallskip

Let us summarize this discussion by providing the list of modes that participate in the path integral, when applying the Hodge decomposition of the metric fluctuations:
\begin{itemize}
    \item $\phi_{\mu\nu}$: All modes participate.
    \item $\eta_\mu$: All modes participate except those in the kernel of $\Delta_1$, which are Killing vectors.
    \item $\chi$: All modes participate except those in the kernel of $ \widetilde{\Delta}_0$, which are non-isometric CKV.
    \item $h$: All modes participate.
\end{itemize}

\subsection{Check: four-dimensional action}

To conclude this section, we briefly recover the four-dimensional results of Volkov and Wipf \cite{Volkov:2000ih} starting from our expressions in arbitrary dimensions and specializing to $D=4$.

\smallskip

First of all, the spin 2 Laplacian $\Delta_2$ has a dimension-independent expression, after simplifying the Ricci tensors using the equations of motion. This is clear from the form on the right-hand side of \eqref{eq:Delta2eee}. The same observation applies to $\Delta_1$ and therefore automatically matches the result of \cite{Volkov:2000ih}.

\smallskip

The only nontrivial sector to check is the scalar one. First, we notice that when $D=4$, the decoupling of $h$ and $\chi$ requires the following relation between $\gamma$ and $\beta$
\beq
\gamma = \frac{1}{2\beta-1}.
\eeq
If we replace $\beta= 2/\beta_{VW}$, where $\beta_{VW}$ is the parameter defined in \cite{Volkov:2000ih}, then we reproduce $\beta_{VW} = 4 \gamma/(1+\gamma)$ which corresponds to equation (3.14) in \cite{Volkov:2000ih}. When $D=4$, our  differential operators acting on scalar fields appearing in the action become
\beq
\Delta_0 = - \nabla^2, ~~~~\widetilde{\Delta}_0 = 3 \Delta_0 - 4 \Lambda,~~~~\widetilde{\Delta}_0^\gamma = \gamma \widetilde{\Delta}_0 - \Delta_0.
\eeq
These match the differential operators that appear in \cite{Volkov:2000ih}. The action in that reference has the same form as ours in terms of these operators and therefore we recover their results. We also note that Equation (3.16) of \cite{Volkov:2000ih} has a typo in the action $h$ and should involve $\widetilde{\Delta}_0^\gamma$ instead of $\widetilde{\Delta}_0$.

\section{General strategy to evaluate the phase}
\label{sec:strategy}

In this section we use the quadratic expansion of the Einstein action given earlier to evaluate the phase of the gravitational path integral. We separately analyze the tensor modes, then the vectors, and finally the scalar modes. We show that all negative tensor modes contribute to the phase, no vector mode does, and there are cancellations in the scalar modes leading to only a reduced number of them potentially generating a phase.

\subsection{Tensor modes}
At short distances compared to the scale set by the cosmological constant, the spectrum of $\Delta_2$ is manifestly positive, since it becomes minus the Laplacian which is a positive operator in flat space. This implies that the $\phi_{\mu\nu}$ action has the right overall sign to be convergent for most modes, up to the possible presence of negative modes at longer distances. If the geometry is compact, the number of potential negative modes of $\Delta_2$ can only be finite. Finally, since all modes in $\phi_{\mu\nu}$ participate in the path integral, we do not need to worry about excluding modes. 

\smallskip

The result is that for each negative mode in $\phi_{\mu\nu}$, we need to rotate the contour to make the integral converge. Let us then denote
\beq
N_2 = \text{\{Number of negative modes of $\Delta_2$\}}.
\eeq
In order to make the path integral over a negative mode convergent we rotate $\phi_{\mu\nu} \to \i \phi_{\mu\nu}$ (without a minus sign). Therefore, the contribution to the phase from the tensor modes is $(-\i)^{N_2}$. We will discuss this contribution in explicit examples later, but let us emphasize that although $S^D$ has no negative tensor modes, other simple geometries such as $S^2 \times S^{D-2}$ do have them, as originally found by Ginsparg and Perry \cite{GINSPARG1983245}, so this is not an exotic possibility.

\subsection{Vector modes}
We show that the vector sector does not present negative modes that could potentially lead to a contribution to the phase. This is valid for any geometry.

\smallskip

The positivity of $\Delta_1$ is implicit in some derivations of the previous section, but it is worth separately emphasizing here. Let us define a pure vector mode 
\beq
h^{(\eta)}_{\mu\nu} = \nabla_\mu \eta_\nu + \nabla_\nu \eta_\mu,~~~\nabla_\rho \eta^\rho = 0.
\eeq
Using the inner product defined in equation \eqref{eq:normvect} we already showed that
\beq\label{eq:VSID}
\int \d^D x\, \sqrt{g} \, h^{(\eta)}_{\mu\nu}\, h^{(\eta)}{}^{\mu\nu}  = 2\int \d^D x\, \sqrt{g} \, \eta_\mu  \Delta_1 \eta^\mu .
\eeq
The left-hand side of this equality is obviously non-negative, since it is the integral of a non-negative quantity. This implies that the spectrum of $\Delta_1$ is also necessarily nonnegative and guarantees that the measure of integration over $\eta$, implicitly defined in \eqref{eq:measuredef}, is real and positive (we discuss zero-modes of $\Delta_1$ shortly). Since the differential operator appearing in the quadratic expansion of the Einstein action is
\beq
\delta^2 I \supset \frac{\gamma}{32\pi G_N} \int \d^D x\, \sqrt{g}\, \eta_\mu  \Delta_1^2 \eta^\mu,
\eeq
there cannot be any negative modes in the vector sector. This can also be seen by noticing that the term in the action is equal to
\beq\label{eq:etactiongft}
\int \d^D x\, \sqrt{g}\, \eta_\mu  \Delta_1^2 \eta^\mu= \int \d^D x \, \sqrt{g} \, (\nabla^\mu h^{(\eta)}_{\mu\nu} )^2\geq 0.
\eeq
This can be derived by integration by parts. An even more direct derivation is the following. Since the vector modes are pure diffeomorphisms, they do not appear in the quadratic expansion of the original Einstein action. Therefore, the gauge-fixed action for the vector modes can arise only from the gauge-fixing term. Since vector modes are traceless, the right-hand side of \eqref{eq:etactiongft} is precisely the gauge fixing term that we introduced in the previous section, which is positive by construction. Therefore the vector modes action should be positive as well.

\smallskip

Finally, we can discuss the zero modes of $\Delta_1$. Equation \eqref{eq:VSID} guarantees, since the integrand in the left-hand side is nonnegative, that if we find a zero-mode of $\Delta_1$ then
\beq
\Delta_1 \eta_\mu = 0,~~~~\Rightarrow~~~~h^{(\eta)}_{\mu\nu}=0.
\eeq
Since these are diffeomorphisms that leave the original metric $g_{\mu\nu}$ invariant, the geometric interpretation is as Killing vector fields as already mentioned in section \ref{sec:PIM}. Since they have an associated metric fluctuation that vanishes, the Killing vectors should be excluded from the path integration over $\eta_\mu$ and all modes that participate in the path integral are strictly positive modes.

\subsection{Scalar modes}

The last sector we need to consider are the scalar modes $\chi$ and $h$. The action is given by
\beq\label{eq:scalaronly}
\delta^2 I \supset \frac{4}{D^2} \langle \chi | \Delta_0 \widetilde{\Delta}_0 \widetilde{\Delta}_0^\gamma \chi \rangle - \frac{D-2}{2\gamma D^2 } \langle h | \widetilde{\Delta}_0^\gamma h\rangle.
\eeq
The first result we will need is the Lichnerowicz-Obata theorem \cite{yano1970integral}, which states that the operator
\beq
\Delta_0 \widetilde{\Delta}_0 \geq 0,
\eeq
is nonnegative. More concretely, the theorem claims that the eigenvalues of the Laplacian are either zero $\Delta_0=0$ or satisfy the bound
\beq
\Delta_0 \geq \frac{2 D \Lambda}{(D-1)(D-2)}=\frac{R}{D-1},
\eeq
which is a straightforward consequence of $\Delta_0 \widetilde{\Delta}_0$ being nonnegative. In other words, the only  negative mode of $\widetilde{\Delta}_0$ is the constant one. The proof is very simple and relies on the inner product appearing in Section \ref{sec:PIM}. Consider a pure $\chi$ mode 
\beq
h^{(\chi)}_{\mu\nu} = 2 \left( \nabla_\mu \nabla_\nu - \frac{1}{D} g_{\mu\nu} \nabla^2\right) \chi.
\eeq
As we did in section \ref{sec:PIM} one can show integrating by parts that
\beq
\int \d^D x \, \sqrt{g} \, h^{(\chi)}_{\mu\nu}\,h^{(\chi)}{}^{\mu\nu} =\frac{4}{D} \int \d^D x \, \sqrt{g} \,\chi \Delta_0 \widetilde{\Delta}_0 \chi.
\eeq
This immediately implies that $\Delta_0 \widetilde{\Delta}_0$ is a nonnegative operator. There are two types of zero modes. We can have constant scalar profiles $\chi_0$ with $\Delta_0 \chi_0=0$. We assume the space is connected and therefore there are no nontrivial solutions of this equation. We can also have modes that annihilate $\widetilde{\Delta}_0$ and we already saw around equation \eqref{eq:SCKV} that those are CKV. We already argued in section \ref{sec:PIM} that the modes in the kernel of $\Delta_0 \widetilde{\Delta}_0$ should be excluded from the $\chi$ path integral. When restricted to modes that appear in the $\chi$ path integral, $\Delta_0 \widetilde{\Delta}_0$ is therefore a strictly positive operator.

\smallskip

 Let us go back to the action \eqref{eq:scalaronly}. We argued in the previous section that at short distances $\widetilde{\Delta}_0^\gamma$ is a positive operator. This implies that on compact spaces all the eigenvalues of $\widetilde{\Delta}_0^\gamma$ are positive, up to a finite number of negative modes. We just saw that the other differential operator that appears $\Delta_0\widetilde{\Delta}_0$ has positive eigenvalues on the integration space for $\chi$. The action \eqref{eq:scalaronly} then suggests that to make the integral convergent we should integrate over real $\chi$ but rotate $h\to \i h'$ such that 
\beq
\delta^2 I \supset \frac{4}{D^2} \langle \chi | \Delta_0 \widetilde{\Delta}_0 \widetilde{\Delta}_0^\gamma \chi \rangle + \frac{D-2}{2\gamma D^2 } \langle h' | \widetilde{\Delta}_0^\gamma h'\rangle.
\eeq
This will make the integral over all but a finite number of modes convergent. The full rotation of a local field (without excluded modes) such as $h$  does not generate any nontrivial phase  \cite{Polchinski:1988ua}. Therefore, after changing variables from $h$ to $h'$, the overall phase of the path integral does not change. 

\smallskip

The final step is to take care of the negative modes of $h'$ and $\chi$, which arise solely from negative modes of $\widetilde{\Delta}_0^\gamma$. The existence of a negative mode implies that we need to rotate the contour of $\chi \to \i \chi'$ to make it convergent, while for $h'$ we need to go back to the original contour $h' \to - \i h$.\footnote{Note that our procedure is inherently gauge independent and consistent with Procedure I in \cite{Ivo:2025yek}.} Therefore, each negative mode of $\chi$ introduces a factor of $(-\i)$, while a negative mode of $h'$ introduces a factor of $(+\i)$.

\smallskip

At this point, an important cancellation takes place. The sign of the action for $\chi$ and $h'$ is dictated only by the sign of $\widetilde{\Delta}_0^\gamma$. Therefore, naively, for any negative mode of $h'$ there is a negative mode of $\chi$ which would lead to a perfect cancellation between their phases. This is not entirely correct. If we have a negative mode of $\widetilde{\Delta}_0^\gamma$ that is excluded from the $\chi$ path integral, then the cancellation will fail since no mode in $h'$ is excluded. Therefore, the contour rotation of all the negative modes of $\chi$ and $h'$ produces the same phase as if we only rotated the negative modes of $h'$ that are not part of the physical $\chi$ spectrum. Since the only modes in the $\chi$ spectrum that are excluded are in the kernel of $\Delta_0\widetilde{\Delta}_0$ the counting is drastically simplified. The resulting scalar contribution to the phase is
\beq
(+\i)^{N_0},~~~~N_0=\{\text{$\#$ of modes with $\widetilde{\Delta}_0^\gamma<0$ and $\Delta_0 \widetilde{\Delta}_0=0$}\}.
\eeq
This can be immediately evaluated for generic geometries, as we will see below. Before doing that, let us emphasize that even though the number of negative modes of $\widetilde{\Delta}^\gamma_0$ might depend on $\gamma$, and therefore on gauge-fixing choices, the number of such negative modes in the kernel of $\Delta_0 \widetilde{\Delta}_0$ is independent of $\gamma$ guaranteeing that the phase of the path integral is gauge invariant\footnote{The gravitational path integral also depends on the choice of measure. This can be characterized by the ``deWitt parameter'' introduced in footnote \ref{footnote}. The ``deWitt parameter'' only rescales the path integral measure of integration over $h$ in \eqref{eq:measure} (since it does not affect traceless modes) and does not modify the phase as long as the inner product derived from it is positive definite.}.

\smallskip

To conclude this discussion, we will show that one can use the Lichnerowicz-Obata theorem to evaluate $N_0$ in general. In a connected spacetime, the only zero-mode of $\Delta_0$ are constant profiles $\chi_0$. We can explicitly check that such a mode is a negative mode of $\widetilde{\Delta}_0^\gamma$ using its definition
\beq
\widetilde{\Delta}_0^\gamma \chi_0 = \left(\gamma \widetilde{\Delta}_0 - \frac{D-2}{2}\Delta_0\right) \chi_0 = - \frac{2\gamma D \Lambda}{D-2} \chi_0.
\eeq
This is therefore always a negative mode, since our gauge fixing parameter $\gamma$ is strictly positive. Moreover, a CKV mode $\chi_c$ such that $\widetilde{\Delta}_0\chi_c=0$ satisfies
\beq
\widetilde{\Delta}_0\,\chi_c=0,~~~\Rightarrow~~~\Delta_0 \,\chi_c = \frac{2D \Lambda}{(D-1)(D-2)}\, \chi_c,
\eeq
and therefore
\beq
\widetilde{\Delta}_0^\gamma \,\chi_c = - \frac{D \Lambda}{D-1} \, \chi_c,
\eeq
and are also negative modes of $\widetilde{\Delta}^\gamma_0$ for all $\gamma>1$.  $N_0$ is then given by the number of  CKV plus one. At this point, we can use yet another classic theorem in Riemannian geometry, the Yano-Nagano theorem \cite{yano1970integral}. In our language, this theorem states that solutions to equation $\widetilde{\Delta}_0\chi_c = 0$ on Einstein spaces do not exist except for the round sphere $S^D$. In that case, an explicit mode expansion reproduced in Section \ref{sec:SD} shows that the equation has $D+1$ solutions. Indeed this is the number of generators of $\SO(D+1,1)/\SO(D+1)$, the space of non-isometric conformal symmetries of the sphere. We conclude
\beq
(+\i)^{N_0} = \begin{cases}
    (+\i),~~~~~~~~~~~~\,\text{for $M\neq S^{D}$},\\
    (+\i)^{D+2},~~~~~~~~~\text{for $M= S^{D}$}.
\end{cases}
\eeq
where $M$ is any $D$-dimensional Einstein space that is simply connected.

\subsection{Assembling all the contributions}

The negative modes of $\Delta_2$ in the tensor sector all contribute to the phase of the path integral, no mode contributes to the phase from the vector sector, and the contribution from scalar negative modes is reduced to the trivial mode or the CKV when they exist. 

\smallskip

Therefore, for a generic Einstein space that is connected and $M\neq S^D$ we get
\beq\label{eq:phaseEM}
Z_M = (-\i)^{N_2- 1} \, \cdot  \, |Z_M|,
\eeq
where $N_2$ is the number of tensor negative modes. In addition to simplifying the quadratic expansion of the Einstein action in arbitrary dimensions, equation \eqref{eq:phaseEM} is the main result of this paper. In the next sections, we will evaluate $N_2$ for some special geometries of interest.  For completeness, we also quote the expression for the absolute value for generic $M$ after simplifying all contributions and including the ghosts
\beq
|Z_M| = \exp{\left(\frac{\Lambda \text{Vol}_M }{4(D-2)\pi G_N} \right)} \,\,|Z^{\text{quant.}}_M|,~~~~|Z^{\text{quant.}}_M|\propto \,\left|\frac{\text{det}'{}^{1/2} \Delta_1}{\text{det}^{1/2}\Delta_2}\right|.
\eeq
The exponential term comes from the on-shell action and $\text{Vol}_M$ is the volume of the space. The second term $|Z^{\text{quant.}}_M|$ arises only from the one-loop determinants. We omit the Killing vectors in the numerator and the exponential is the classical Einstein action in spacetime $M$. The functional determinant of $\widetilde{\Delta}_0^\gamma$ in the absolute value cancels between the ghosts, the $\chi$ mode, and the $h$ mode. This was observed in \cite{Gross:1982cv} although the phase was not correctly incorporated. The normalization of $|Z^{\text{quant.}}_M|$ can be evaluated more explicitly. This includes several subtleties regarding logarithmic corrections worked out in \cite{Anninos:2020hfj,Law:2020cpj} that we will not need for the purposes in this paper.

\smallskip

Going back to a generic Einstein space $M$, how should we interpret the presence of zero-modes of $\Delta_2$? Tensor modes in the kernel of $\Delta_2$ are called infinitesimal Einstein deformations. They correspond to infinitesimal directions in the moduli space of Einstein structures, the space of Einstein metrics modulo diffeomorphisms and volume rescalings \cite{koiso1982rigidity}. We should therefore extract these modes from the functional determinant in $|Z_M|$ and interpret the result as a measure of integration over this moduli space.

\section{Application to specific geometries}\label{sec:Applications}

In this section we analyze the spectrum of $\Delta_2$ on some special Einstein spaces $M$, the differential operator acting on TT modes of the metric. Identifying the presence of negative modes will allow us to evaluate the phase of the gravitational path integral on those geometries.

\subsection{de Sitter and $S^D$}
\label{sec:SD}
The simplest Einstein space is the round sphere $S^{D}$ of radius ${\sf R}$ arising from a Wick rotation of Lorentzian dS. The dS metric is given, in global coordinates, by
\beq
\d s^2 = {\sf R}^2 \left( -\d t^2 + \cosh^2 t \, \d \Omega_{D-1}^2\right),
\eeq
where $\d \Omega_{D-1}^2$ is the line element of a unit $D-1$ dimensional sphere, and Einstein's equations set
\beq
\frac{1}{{\sf R}^2} =\frac{2 \Lambda}{(D-2)(D-1)} . 
\eeq
The Euclidean section, obtained by taking $t=-\i \tau$ with real $\tau$ restricted to $-\pi/2\leq \tau \leq \pi/2$, corresponds to the $D$ dimensional sphere $S^D$. A reasonable interpretation of the gravitational path integral on this geometry is that it evaluates the leading contribution to the norm of the Hartle-Hawking wavefunction \cite{Hartle:1983ai}. The phase of the gravitational path integral on $S^D$ was evaluated in \cite{Polchinski:1988ua}, see also \cite{Maldacena:2024spf}, and found to be complex. The ultimate interpretation of this result is still open. In this section we use our formalism to reproduce that result. 

\smallskip

We begin by recalling some facts about the spherical harmonics on a sphere $S^D$ of radius ${\sf R}$. We will only need the scalar, vector, and tensor ones:

\paragraph{Scalar harmonics} We can describe $S^D$ as a sphere $X^A X^A = {\sf R}^2$ inside flat space $\mathbb{R}^{D+1}$, parametrized by Cartesian coordinates $X^A$, $A=1,\ldots, D+1$. The sphere can be parametrized by $X^A = {\sf R}\, \Omega^A$, where $\Omega^2=1$. The scalar harmonics are homogeneous polynomials in the Cartesian coordinates
\beq
Y(\ell) = \frac{1}{{\sf R}^\ell}\, T_{A_1\ldots A_\ell} \,X^{A_1} =T_{A_1\ldots A_\ell} \,\Omega^{A_1}  \ldots \Omega^{A_\ell},~~~~~\ell=0,1,2,\ldots,
\eeq
where $T_{A_1\ldots A_\ell}$ is a tensor that is constant in Cartesian coordinates. It is automatically totally symmetric, and we impose trace-free conditions in all pairs of indices so different $Y(\ell)$ are independent. These functions can be further decomposed into a basis of eigenstates of a choice of Cartan generators of $\SO(D+1)$, but we will not need to do this here. We can show
\beq
\Delta_0 Y(\ell) = \frac{ \ell(\ell+D-1)}{{\sf R}^2}  \, Y(\ell),
\eeq
by writing the scalar Laplacian in $\mathbb{R}^{D+1}$, going to spherical coordinates, and decomposing it into an angular Laplacian and a radial component \cite{Gibbons:1978ji,Rubin:1984tc}. 

\paragraph{Vector harmonics} We can construct vector harmonics in a similar fashion
\beq
Y_{\tilde{A}} (\ell) = \frac{1}{{\sf R}^{\ell}}\, T_{\tilde{A} B A_1 \ldots A_{\ell-1}} X^B X^A_1 \ldots X^{A_{\ell-1}},~~~~\ell=1,2,\ldots,
\eeq
where $\tilde{A}$ denotes an index that is tangent to the sphere, allowing it to be pulled-back to a (dual) vector $Y_{\mu}(\ell)$ on the sphere. The tensor $T$ is again constant in Cartesian coordinates and has the following properties. It is symmetric in all $A_1,\ldots,A_{\ell-1}$, it is antisymmetric with respect to the exchange of $\tilde{A}$ with $B$, and it is traceless under contraction of any pair of indices. A decomposition of the vector Laplacian on radial coordinates shows that, if these conditions hold, then $Y_{\tilde{A}}(\ell)$ is divergence-free and satisfies  
\beq
- \nabla^2 \, Y_{\tilde{A}}(\ell) = \frac{\ell(\ell+D-1) -1}{{\sf R}^2} \,Y_{\tilde{A}}(\ell).
\eeq
Using the fact that the Ricci scalar on any classical solution is proportional to the metric, we can derive
\beq
\Delta_1 Y_{\tilde{A}} (\ell) =\left(-\nabla^2-\frac{2\Lambda}{D-2} \right)\,Y_{\tilde{A}} (\ell) =\frac{(\ell-1)(\ell+D)}{{\sf R}^2} \, Y_{\tilde{A}} (\ell),
\eeq
where $\Delta_1$ is the vector Laplacian defined as in equation \eqref{eq:Delta1eee}. 

\paragraph{Tensor harmonics} The TT tensor harmonics are given by 
\beq
Y_{\tilde{A}_1 \tilde{A}_2} (\ell) = \frac{1}{{\sf R}^{\ell}}\, T_{\tilde{A}_1 B_1 \tilde{A}_2 B_2 A_1 \ldots A_{\ell-1}} X^{B_1} X^{B_2} X^A_1 \ldots X^{A_{\ell-2}},~~~~\ell=2,3,\ldots.
\eeq
The directions $\tilde{A}_i$ are tangential to the sphere and therefore we can pull-back to a tensor $Y_{\mu\nu}(\ell)$ on the sphere. The tensor $T$ is constant in Cartesian coordinates, symmetric in $A_1,\ldots,A_n$, antisymmetric under interchange of $\tilde{A}_i$ with $B_i$, traceless under all contractions, and symmetric under exchange of any pair $\tilde{A}_i B_i$. These conditions imply that the tensor harmonic is TT and
\beq
-\nabla^2 \, Y_{\tilde{A}_1 \tilde{A}_2} (\ell) = \frac{\ell(\ell+D-1)-2}{{\sf R}^2}\, Y_{\tilde{A}_1 \tilde{A}_2} (\ell).
\eeq
More generally the eigenvalues of a tensor with $n$ indices are $(\ell(\ell+D-1)-n)/{\sf R}^2$, see \cite{Rubin:1984tc} for a derivation. We can use this result to derive
\beq
\Delta_2 \,Y_{\tilde{A}_1 \tilde{A}_2} (\ell) = \frac{\ell(\ell+D-1)}{{\sf R}^2} \, Y_{\tilde{A}_1 \tilde{A}_2} (\ell).\label{eq:TSHE}
\eeq
The differential operator $\Delta_2$ was defined in \eqref{eq:Delta2eee} and appears in the Einstein action. To derive this we used that the Riemann tensor of a sphere is a properly symmetrized sum of products of metric tensors. 

\smallskip

We are now ready to reproduce the result of \cite{Polchinski:1988ua}. Let us begin considering tensor harmonics. We see from equation \eqref{eq:TSHE} that all the eigenvalues of $\Delta_2$ are strictly positive since the lowest possible eigenvalue is 
\beq
\Delta_2 \geq \frac{2(D+1)}{{\sf R}^2}.
\eeq
In the notation of section \ref{sec:strategy}, we find that the parameter $N_2=0$ vanishes, and there is no phase arising from the tensor modes. At this point, according to our formalism, we can immediately conclude that the phase of the path integral is
\beq\label{eq:POLRES}
Z_{S^D} = (+\i)^{D+2} \cdot |Z_{S^D}|.
\eeq
The Yano-Nagano theorem guarantees a scalar contribution coming both from the constant mode as well as the $D+1$ CKV of the sphere. In the rest of this section we verify this result by examining the vector and scalar harmonics explicitly.

\smallskip

We gave a general derivation in Section \ref{sec:strategy} that vector modes do not contribute to the phase. The operator $\Delta_1$ is always nonnegative and the zero modes correspond to the Killing vectors of the background. In our case, we can immediately verify that $\Delta_1 =0 $ when $\ell=1$ and those modes are Killing vectors
\beq
\nabla_\mu \eta_\nu + \nabla_\nu \eta_\mu = 0,~~~~\eta_\mu = Y_\mu(1).
\eeq
They are parametrized by an antisymmetric $D+1$ dimensional tensor with two indices, implying there are $D(D+1)/2$ such modes. This is the number of generators of the sphere isometry group $\SO(D+1)$. We remind the reader that these modes do not participate in the path integral over metric fluctuations, and the modes that are excluded from the $\eta_\mu$ path integral are precisely $\ell=1$.

\smallskip

Finally, we consider the scalar sector. According to our rules, we only need to consider the negative modes of $\widetilde{\Delta}_0^\gamma$ restricted to the kernel of $\Delta_0 \widetilde{\Delta}_0$. Modes annihilated by $\Delta_0 \widetilde{\Delta}_0$ are either
\beq
\Delta_0  \,Y(\ell) = \frac{\ell(\ell+D-1)}{{\sf R}^2} \,Y(\ell) = 0,~~~\Rightarrow~~~\ell=0,
\eeq
which is the constant mode present in all spaces, or
\beq
\widetilde{\Delta}_0 \, Y(\ell) = \frac{(\ell-1)(\ell+D)(D-1)}{{\sf R}^2} \,Y(\ell) = 0,~~~\Rightarrow~~~\ell=1,
\eeq
which corresponds to the  CKV that according to Yano-Nagano only exist on the sphere \cite{yano1970integral}. Since $Y(1) \sim T_A X^A$ there are $D+1$ such harmonics and together with the isometries make up $(D+1)(D+2)/2$ generators, the same as the conformal group $\SO(D+1,1)$. We already argued that any mode that is in the kernel of $\Delta_0$ or $\widetilde{\Delta}_0$ will be unavoidably a negative mode of $\widetilde{\Delta}_0^\gamma$ and therefore $N_0=1+(D+1)=D+2$. 

\smallskip 

Including the on-shell action, the sphere path integral is
\beq
Z_{S^D}= \i^{D+2}\,\cdot\,e^{{\sf S}_{\text{dS}}}\, \cdot  \,|Z^{\text{quant.}}_{S^D}|,
\eeq
where 
\beq\label{eq:DSENTROPY}
{\sf S}_{\text{dS}} = \frac{(D-1){\sf R}^{D-2}\Omega_D}{8\pi G_N},
\eeq
is the dS entropy and $\Omega_D = 2\pi^{(D+1)/2}/\Gamma((D+1)/2)$ is the area of the $D$ sphere. The term $|Z^{\text{quant.}}_{S^D}|$ was calculated explicitly in \cite{Volkov:2000ih, Anninos:2020hfj} but will not play an important role in our analysis, see Appendix B of \cite{Maldacena:2024spf} for a summary of the explicit form of this contribution.

\subsection{Black holes in de Sitter and $S^2 \times S^{D-2}$}\label{sec:Nariai}

Let us consider a massive black hole in dS. This geometry is described by the Schwarzschild-dS metric
\beq
\d s^2 = {\sf R}^2\left(- f \d t^2 + \frac{\d r^2}{f} + r^2 \, \d \Omega_{D-2}^2\right),~~~~~f=1-r^2 - \frac{\mu}{r^{D-3}},
\eeq
where ${\sf R}$ is the radius of dS and $\mu$ is proportional to the black hole mass. For generic values of the mass this geometry has two horizons at $f=0$. The largest root of $f$ corresponds to the cosmological horizon and the second largest to the black hole horizon. They have different surface gravity and therefore different temperatures. For this reason, the Euclidean continuation of the geometry is not smooth in the usual fashion. Moreover, even in Lorentzian signature, a naked singularity can develop if the black hole mass is too large. The maximum value for the mass is given by
\beq
\mu_{N} = \frac{2}{D-3} r_{N}^{D-1},~~~~~~r_N = \sqrt{\frac{D-3}{D-1}},
\eeq
and it is called the Nariai limit. As the mass approaches $\mu_N$ the two horizons become extremal with a vanishing temperature, and the size of the spatial $S^{D-2}$ becomes constant everywhere between the two horizons. In Euclidean signature $t=-\i \tau$ and after a change of variables
\beq
\mu = \mu_N + (D-1)r_N^{D-1} \epsilon^2,~~~~r= r_N(1+\epsilon \cos \rho) ,~~~~~\tau = \frac{\eta}{\epsilon(D-1) r_N},
\eeq
where we take $\epsilon \to 0$ to approach the Nariai geometry, the metric becomes the product of a two-sphere and a $D-2$ dimensional sphere
\beq\label{eq:NariaiM}
\d s^2 = {\sf R}_1^2 \,\, \underbrace{\left( \sin^2 \rho \,\, \d \eta^2 + \d \rho^2 \right)}_{S^2} + {\sf R}_2^2\,\, \d \Omega_{D-2}^2,
\eeq
with the radii of the two spheres given by
\beq\label{eq:radiiS2SD}
{\sf R}_1^2 = {\sf R}^2\frac{1}{D-1} ,~~~~{\sf R}_2^2 = {\sf R}^2\frac{D-3}{D-1}.
\eeq
Notice that only in $D=4$ the two radii are equal. The first factor in the metric is the Euclidean continuation of the static patch geometry, and the two poles of the sphere $\rho=0$ and $\pi$ correspond to the black hole and cosmological horizons in Lorentzian signature. We could have also found this solution by taking \eqref{eq:NariaiM} as an ansatz and verifying that it solves Einstein's equation only when the radii take these values. 

\smallskip

There are two possible interpretations of the $S^2 \times S^{D-2}$ geometry and each implies a different phase for $Z_{S^2 \times S^{D-2}}$. One interpretation is that it calculates the norm of the Hartle-Hawking wavefunction, conditional on the future spatial topology being $S^1 \times S^2$, see \cite{Turiaci:2025xwi} for a recent account and references therein. This would make it a non-perturbative small  correction to the $S^D$ path integral and requires the path integral on $S^2 \times S^{D-2}$ to be real. Another possible interpretation, put forth in \cite{GINSPARG1983245} and \cite{Bousso:1995cc,Bousso:1996au}, is that the path integral on $S^2 \times S^{D-2}$ computes the production rate of black holes in dS. This interpretation requires the path integral to be imaginary instead. In this section, we will be agnostic to the interpretation and simply focus on the technical problem of evaluating the phase of the gravitational path integral $Z_{S^2 \times S^{D-2}}$, which was not considered previously.

\smallskip

We will apply the general strategy derived in Section \ref{sec:strategy} and explicitly verify it with the sphere harmonics. According to the Yano-Nagano theorem, there should be effectively a single scalar mode that contributes to the phase, the constant mode. To verify this claim explicitly for $S^2 \times S^{D-2}$, we can expand the most general function in a product of $S^2$ harmonics with mode number $\ell_1$ and a $S^{D-2}$ harmonic with mode number $\ell_2$. A simple calculation gives the eigenvalue of the scalar Laplacian
\beq
\Delta_0 \, Y(\ell_1) Y(\ell_2) = \frac{2\Lambda}{D-2} \sum_{j=1,2} \frac{\ell_j (\ell_j +D_j-1)}{D_j-1}\,  Y(\ell_1) Y(\ell_2),
\eeq
where $D_1 = 2$ and $D_2 = D-2$. The constant mode with $\ell_1=\ell_2=0$ is again in the kernel of $\Delta_0\widetilde{\Delta}_0$ and therefore contributes to the phase. A harmonic annihilated by $\widetilde{\Delta}_0$ would need to satisfy
\beq
\widetilde{\Delta}_0 \,  Y(\ell_1) Y(\ell_2) = \left(\frac{2(D-1)\Lambda}{D-2} \sum_{j=1,2} \frac{\ell_j (\ell_j +D_j-1)}{D_j-1}-\frac{2 D \Lambda}{D-2} \right)\, Y(\ell_1) Y(\ell_2) = 0,
\eeq
or equivalently
\beq
 \ell_1 (\ell_1 +1)+ \frac{\ell_2 (\ell_2 +D-3)}{D-3} = \frac{ D}{D-1}
\eeq
This equation has no positive integral solutions for $\ell_1$ and $\ell_2$, and therefore the kernel of $\widetilde{\Delta}_0$ is empty. This confirms that, other than the sphere, the only scalar mode that contributes effectively to the phase is the constant one.

\smallskip

As proven in Section \ref{sec:strategy} there are no negative modes and no phase arising from the vector sector. The only thing left to do is to compute $N_2$, the number of negative modes of $\Delta_2$. In the next section and in Appendix \ref{app:delta2spectrum} we will give a general analysis that applies to a product of an arbitrary number of spheres with arbitrary dimension. In this section, we will simply quote the form of the lowest lying eigenmodes of $\Delta_2$. 

\smallskip

The lowest eigenvalue of $\Delta_2$ on a $S^{D_1} \times S^{D_2} \times \ldots$  always arises from pure traces on each individual sphere that are traceless in the full metric, as explained in Appendix \ref{app:delta2spectrum}. Moreover we can minimize $\Delta_2$ by taking each subtrace to be constant. Let us denote $\mu_1,\nu_1=\rho,\eta$ and $\mu_2,\nu_2$ runs over the $D-2$ coordinates of $S^{D-2}$. The lowest mode on $S^2 \times S^{D-2}$ is proportional to
    \beq\label{eq:YLS2SD}
  h_{\mu\nu}= \, \begin{pmatrix}
        (D-2)g_{\mu_1\nu_1} & 0  \\
        0 &  - 2g_{\mu_2\nu_2}
   \end{pmatrix}
    \eeq
    where we fixed the relative coefficient between the $S^2$ and $S^{D-2}$ blocks to cancel the full trace. This mode is also transverse. The eigenvalue of $\Delta_2$ is
    \beq
\Delta_2 \, h_{\mu\nu} =- \frac{4 \Lambda}{D-2}\, h_{\mu\nu},
    \eeq
    making it a negative mode. This instability of $S^2 \times S^{D-2}$ was first discovered by Ginsparg and Perry \cite{GINSPARG1983245} in $D=4$ and extended to $D$ dimensions in \cite{Gonzalez-Diaz:1987rxk}. We will show in the next section that this is the only negative mode on $S^2 \times S^{D-2}$. The idea is the following. Naively, the first excited mode of $\Delta_2$ takes the same form as \eqref{eq:YLS2SD} but multiplied by $Y(\ell_1=0)Y(\ell_2=1)$ which would also have a negative eigenvalue. Nevertheless, one can show that this mode cannot be completed into a transverse tensor. A mode with $Y(\ell_1=1)Y(\ell_2=0)$ would be a zero-mode but again cannot be made transverse. All other modes one can build that do not have this structure can be shown to be positive, which we do in a more general setting in Appendix \ref{app:delta2spectrum}.

\smallskip

Since we found a single negative tensor mode $N_2=1$ and the phase of the path integral is dimension-independent and real. The volume of the space is $\text{Vol}_{S^2 \times S^{D-2}}= {\sf R}_1^2 \, \Omega_2 \cdot {\sf R}_2^{D-2}\,\Omega_{D-2}$ and therefore the full gravitational path integral is given by
\bea
Z_{S^2 \times S^{D-2}} = \exp{\left( 2\left(\frac{D-3}{D-1}\right)^{\frac{D-2}{2}} {\sf S}_{\text{dS}}\right)}\, \cdot \, \Big|Z^{\text{quant.}}_{S^2 \times S^{D-2}}\Big| ,
\ea
where we defined ${\sf S}_{\text{dS}}$ as the dS entropy in \eqref{eq:DSENTROPY}. Notice that for any dimension $D>3$ the exponential term of the Nariai solution is always smaller than the contribution of the sphere. The absolute value $|Z^{\text{quant.}}_{S^2 \times S^{D-2}}|$ was evaluated in $D=4$ \cite{Volkov:2000ih} but not for $D\geq 5$ as far as we know.

\smallskip

In \cite{GINSPARG1983245,Volkov:2000ih} it is incorrectly stated that the path integral $Z_{S^2 \times S^{D-2}}$ is imaginary due to the tensor negative mode. The mistake is that those authors did not include the scalar sector in the considerations of the phase. Our result of a real path integral therefore cannot be interpreted by itself as computing the decay rate of dS, as originally intended by Ginsparg and Perry in \cite{GINSPARG1983245}.  We will comment on the interpretation of this result in the concluding section \ref{sec:Discussions}, after analyzing more cases.

\subsection{Generalization to $S^{D_1}\times \ldots \times S^{D_n}$}\label{sec:PRODSPHERE}

We evaluate the phase of the gravitational path integral on a product of an arbitrary number of spheres $S^{D_1}\times \ldots \times S^{D_n}$ with total dimension $D=\sum_j D_j$. We are not aware of any Lorentzian metric that would lead to this geometry in an extremal limit, unless perhaps one of the $D_j$ is equal to two. Instead, we can simply postulate as an ansatz the metric
\beq
\d s^2 = \sum_j\,\, {\sf R}_j^2 \,\cdot\, \d \Omega_{D_j}^2.
\label{eq:metricSDi}
\eeq
One can show that this geometry solves Einstein's equations with cosmological constant $\Lambda$ as long as
\beq
\frac{R_j}{D_j} = \frac{R}{D} = \frac{2 \Lambda}{D-2},
\eeq
where $R_j$ are the individual Ricci scalars, and from this expression we can extract the individual radii of curvature
\beq
\frac{1}{{\sf R}_j^2} = \frac{2\Lambda}{D-2} \frac{1}{D_j-1},~~~\text{or}~~~{\sf R}_j^2 = {\sf R}^2\, \frac{D_j-1}{D-1}.
\eeq
The second expression matches what we found in the $S^2 \times S^{D-2}$ case in equation \eqref{eq:radiiS2SD}. The on-shell action of this geometry is
\beq
Z_M =\exp{\left( \frac{\prod_j(D_j-1)^{D_j/2} \Omega_{D_j}}{(D-1)^{D/2}\Omega_D} \, \cdot \,{\sf S}_{\text{dS}} \right)} \, \cdot \, Z^{\text{quant}}_M,
\eeq
where again we wrote it in terms of the dS entropy. (One can verify that the term in the exponential is always smaller than one.) In this section, we evaluate the phase of $Z^{\text{quant}}_M$ and therefore the overall phase of the gravitational path integral on these spaces. Unlike $S^2 \times S^{D-2}$, we will find that in this more general case the answer depends on the number of spheres and dimensionality. This path integral could either be interpreted as a correction to the norm of the Hartle-Hawking wavefunction or the decay rate of dS, but we will not commit to it in this discussion and simply focus on the evaluation of the phase.

\smallskip

In Appendix \ref{app:delta2spectrum} we present a way to derive the spectrum of $\Delta_2$ acting on TT tensor for any product of arbitrary Einstein manifolds. The formalism is based on \cite{Yasuda:1984py} although we correct some mistakes present in that reference. The upshot of that analysis is that negative modes of $\Delta_2$ can only originate from the trace components on each $S^{D_i}$, located in the diagonal blocks of the full tensor $$\begin{pmatrix}
  \frac{1}{D_1}g_{\mu_1\nu_1} h_1 & \cdots & 0\\
\vdots & \ddots & \vdots\\
0 & \cdots & \frac{1}{D_n} g_{\mu_n\nu_n} h_n.\\
\end{pmatrix}$$ with the following eigenvalues and eigenfunctions
\ie
h_i=\prod_{j}Y_j(\ell_j)~~~&i=2,\dots,n, ~~~~h_1=-\sum_{i=2}^n h_i
\\
\left(\sum_{j}\Delta_0^{(j)}-\frac{4\Lambda}{D-2}\right)h_i&=\frac{2\Lambda}{D-2}\left(\sum_j\frac{\ell_j(\ell_j+D_j-1)}{D_j-1}-2\right)h_i.
\fe
We will identify the set of $\{\ell_j\}$ that lead to negative modes and evaluate $N_2$ and leave Appendix \ref{app:delta2spectrum} for an explanation of why considering this sector is enough. The first several modes are given by
\begin{itemize}
\item 
The constant trace modes with $\ell_j=0$ on all the subspaces, constrained by the traceless condition of the full tensor, namely
\ie
\Delta_2 \begin{pmatrix}
 c_1 g_{\mu_1\nu_1} & \cdots & 0\\
\vdots & \ddots & \vdots\\
0 & \cdots & c_n g_{\mu_n\nu_n}\\
\end{pmatrix}=-\frac{4\Lambda}{D-2}\begin{pmatrix}
 c_1 g_{\mu_1\nu_1} & \cdots & 0\\
\vdots & \ddots & \vdots\\
0 & \cdots & c_n g_{\mu_n\nu_n}\\
\end{pmatrix}\quad\text{with }\sum_i c_i D_i=0
\fe
In total there are $n-1$ constant modes, and they are all negative modes. When $n=2$ and $D_1=2$, these reduce to the negative modes found in \cite{Ginsparg:1993is,Gonzalez-Diaz:1987rxk}.

\item The next case is when $\ell_j=1$ and $\ell_k=0$ for all other $k\neq j$, namely
\beq
h_i = c_i Y_j(1),~~~i=1,\ldots,n.
\eeq
Their eigenvalues are given by $-\frac{2\Lambda}{D-2}\frac{D_j-2}{D_j-1}$. This is a negative mode for $D_j>2$ and a zero mode if $D_j=2$. However, not all of them are TT. \footnote{In an earlier version of the paper, we had not realized that these modes carry a negative $\Delta_2$ eigenvalue. We thank Victor Ivo for pointing it out to us. These modes were also incorrectly excluded in \cite{Yasuda:1984py}.}  Transversality reduces to $\nabla_{\mu_i} h_i = 0$. The modes along $S^{D_k}$ with $k\neq j$ immediately satisfy this constraint since $h_k$ is independent of $S^{D_k}$. This fails when we consider $h_j$ since $\nabla_{\mu_j} h_j = c_j \nabla_{\mu_j}  Y_j(1)\neq 0$, but we can restore transversality by setting $c_j=0$. The TT mode has $n-1$  coefficients $c_k$ with $k\neq j$ satisfying one constraint $\sum_{k\neq j} c_k=0$, leaving $n-2$ free parameters. For each choice of the $c_i$ we have $D_j+1$ independent choices of $Y_j(1)$. The total degeneracy is given by 
\ie
(n-2)\sum_{i ~\text{for}~D_i>2}(D_i+1),
\fe
where we removed the modes corresponding to a sphere $D_i=2$ since they are zero modes and should not be rotated. This means that the product of more than two spheres involving an $S^2$ presents infinitesimal Einstein deformations.

\item One can easily show that all the other modes are positive, see Appendix \ref{app:delta2spectrum} for a complete analysis in a more general setting. To illustrate this in a simple example within the diagonal sector, having two scalar harmonics  turned on $\ell_j\geq 1$ and $\ell_k\geq1$ with all other $\ell_{i\neq k,j}=0$ already produces a positive mode. Considering only one $\ell_j=2$ with $\ell_{k\neq j}=0$ also produces a positive mode.\footnote{These modes are not themselves transverse. As explained around equation \eqref{eq:transversality}, one can combine them with other scalar modes to be made transverse. This is not important for our discussion here since they will never produce negative modes.} This is a generate feature consequence of the Lichnerowicz-Obata theorem valid beyond spheres. There are also TT modes on $S^{D_1}\times \ldots \times S^{D_n}$ made out of TT tensors or divergenceless vectors on individual or pairs of spheres, but they all have a positive spectrum.

\end{itemize}

In total, the number of tensor negative modes is given by 
\ie
N_2=n-1+(n-2)\sum_{i ~\text{for}~D_i>2} (D_i+1)=n-1+(n-2)(D+n-3n_2),
\fe
where $n_2$ is the number of $S^2$ factors in the total geometry. The phase of the path integral on $M=S^{D_1} \times \ldots \times S^{D_n}$ is 
\bea
Z_M =(-\i)^{(n-2)(D+n+1-3n_2)}\, \cdot \, \exp{\left( \frac{\prod_j(D_j-1)^{D_j/2} \Omega_{D_j}}{(D-1)^{D/2}\Omega_D} \, \cdot \,{\sf S}_{\text{dS}} \right)} \, \cdot \,| Z^{\text{quant}}_M|.
\ea
This answer reproduces known results in some special cases. If the manifold is made of a single sphere $n=1$ with $D>2$  so $n_2=0$ the phase becomes
 $(n-2)(D+n+1-3n_2) = - D-2$ and reproduces the answer in \cite{Polchinski:1988ua,Maldacena:2024spf}.
When we consider a product with only two factors $M=S^{D_1} \times S^{D_2}$ with $n=2$ the partition function is real and positive. When $D_1=2$ this reproduces and generalizes the result we found in the previous section regarding the Nariai geometry.

\subsection{Results for other Einstein spaces}

Although we reduced the evaluation of the path integral phase to analyzing only the spectrum of $\Delta_2$ acting on TT modes, this is the most complicated sector in general. In this section, we collect some results that we were able to derive that are valid for a large class of Einstein spaces. 

\subsubsection{Rigid stable Einstein spaces}

There is a special class of geometries called rigid Einstein spaces. These are solutions of Einstein's equation with a spectrum of $\Delta_2$ that has no zero modes, see \cite{koiso1980rigidity}. These are isolated points in moduli space, that would otherwise be parameterized by zero modes of $\Delta_2$. In this paper, we restrict the definition of rigid Einstein space to those where $\Delta_2$ is strictly positive so that they are also stable. The simplest example we already encountered is the sphere. This implies that the phase of the path integral over such spaces is either $\i^{D+2}$ (for the sphere) or $\i$ (for any other manifold that is rigid and is not the sphere). A list of rigid spaces can be found in \cite{koiso1980rigidity} or in Appendix A of \cite{Bonifacio:2020xoc}. Combining the classical action, given by the volume taken from \cite{Bonifacio:2020xoc} with the phase we computed, we can already write down the answer for these rigid manifolds:
\begin{itemize}
    \item  The first example in the list is $S^D$. We already discussed this case, but we repeat the answer for comparison
    \beq
Z_{S^D}= \i^{D+2}\, \cdot \, \exp{\left({\sf S}_{\text{dS}}\right)} \, \cdot \, \big| Z^{\text{quant.}}_{S^D}\big|.
    \eeq
    \item The second example is the projective space $\mathbb{RP}^D$ which is a quotient $S^D/\mathbb{Z}_2$. The gravitational path integral is 
    \beq
Z_{\mathbb{RP}^D} = \i\, \cdot \, \exp{\left(\frac{1}{2}\,{\sf S}_{\text{dS}}\right)} \, \cdot \, \big| Z^{\text{quant.}}_{\mathbb{RP}^D}\big|.
    \eeq
    The harmonics on this space are the same as in the sphere but with restrictions. The scalar and tensor harmonics only include even $\ell$ (thus eliminating the CKV, consistent with the Yano-Nagano theorem), while the vectors only include odd $\ell$.
    \item The third example is the complex projective space $\mathbb{CP}^n$ with $D=2n$ and the Fubini-Study metric. The path integral is
    \beq
Z_{\mathbb{CP}^{n}}= \i \, \cdot \, \exp{\left(\frac{(\frac{2n+2}{2n-1})^{n}\Gamma(n+\frac{1}{2})}{2\sqrt{\pi}\, \Gamma(n+1)}\,{\sf S}_{\text{dS}}\right)}\, \cdot \, | Z^{\text{quant.}}_{\mathbb{CP}^{n}}|.
    \eeq
    This space can also be interpreted as $S^{2n+1}/S^1$ with the quotient obtained from the Hopf fibration. The mode expansion can then also be inferred from the sphere harmonics. The case $n=2$ matches with the answer in \cite{Gibbons:1978zy}.
    \item The fourth example is the quaternionic projective space $\mathbb{HP}^n$ with its standard metric and dimension $D=4n$. It can arise by an $S^3$ quotient of $S^{D+3}$ using a quaternionic Hopf fibration. The path integral is
    \beq
    Z_{\mathbb{HP}^{n}}= \i \, \cdot \, \exp{\left(\frac{(\frac{4n+8}{4n-1})^{2n}\Gamma(\frac{4n+1}{2})}{2\sqrt{\pi}\Gamma(2n+2)}\,{\sf S}_{\text{dS}}\right)}\, \cdot \, \big| Z^{\text{quant.}}_{\mathbb{HP}^{n}}\big|
    \eeq
    \item The last example in the list is the Cayley plane $\mathbb{OP}^2$, with its standard metric, which is a $D=16$ dimensional space. The path integral is
    \beq
    Z_{\mathbb{OP}^{2}}= \i \, \cdot \, \exp{\left(\frac{255879}{390625}\,{\sf S}_{\text{dS}}\right)}\, \cdot \, \big| Z^{\text{quant.}}_{\mathbb{OP}^{2}}\big|
    \eeq
    
\end{itemize}
In all cases, we compared the action with that of the sphere of the same dimensionality and find that they are all subleading saddles. This is consistent with the expectation that the spheres have the largest volume among classical solutions to the Einstein equations, as proved in \cite{bishop2011geometry}, p. 254. Although these are exotic examples, the discussion is intended mostly to illustrate the application of the formalism in Section \ref{sec:strategy}. The sphere and projective space are clearly familiar as dS and an orbifold thereof. It is shown in \cite{Page:2009dm} that starting from the Taub-NUT dS geometry, one can take an extremal limit that leads to the complex projective space $\mathbb{CP}^2$. We have already concluded that the phase of this contribution is $\i$. The action is down, compared to dS, by a factor of $3/4 \cdot {\sf S}_{\text{dS}}$ in four dimensions, but dominates over the Nariai action, which is $2/3 \cdot {\sf S}_{\text{dS}}$. To our knowledge, the quaternionic projective space and the Cayley plane do not have any physical interpretation.

\subsubsection{Product of Einstein spaces $M_1\times \ldots \times M_n$}
In this section we consider what happens when we generalize the calculation in Section \ref{sec:PRODSPHERE} to an Einstein space $M$ formed by a product
\beq
M_1 \times \ldots \times M_n,
\eeq
where $M_j$ are Einstein spaces of dimension $D_j$. The full geometry solves Einstein's equation if
\beq
\frac{R_j}{D_j} = \frac{R}{D},
\eeq
for all $j=1,\ldots,n$. The partition function on these spaces is given by
\beq
Z_M = \exp{\left( \frac{\text{Vol}_M}{{\sf R}^{D} \Omega_D}\, {\sf S}_{\text{dS}}\right)}\,\cdot \, Z_M^{\text{quant}}.
\eeq
We will evaluate here the phase of $Z_M^{\text{quant}}$ in terms of properties of each factor $M_j$. 

\smallskip

Let us count the number of unstable TT modes on $M$. If each of the factors $M_j$ have themselves negative eigentensors of $\Delta_2$, those will also be present in the spectrum of $\Delta_2$ acting on $M$. We will denote by $N_2^{(j)}$ the number of negative eigentensors of $M_j$. What other negative modes can we have? Based on Theorem 3.3.1 of \cite{Yasuda:1984py}, we can universally claim that there will always be at least $n-1$. Their form is the following. Decompose the coordinates on $M$ into coordinates $\mu_i,\nu_i=1\ldots,D_i$ on individual factors $M_i$, and consider the ansatz
\bea\label{eq:YASUDANM}
h_{\mu\nu}=\begin{pmatrix}
\ddots &  &  &  0\\
 & g_{\mu_i\nu_i}\frac{1}{D_i} h_i &  \\
 0& & &\ddots\\
\end{pmatrix},~~~\sum_i h_i =0.
\ea
If the scalar modes $h_j$ are all constant on $M$ and satisfy the constraint $\sum_j h_j =0$, we obtain
\ie
\,&\Delta_2 h_{\mu\nu} = - \frac{4\Lambda}{D-2} \, h_{\mu\nu}.
\fe
These fluctuations are determined by $n-1$ free parameters.

\smallskip

According to the analysis in Appendix \ref{app:delta2spectrum}, there is only one other source of negative modes. Consider the same ansatz as in equation \eqref{eq:YASUDANM}, but where the $h_i$ are given by a single excited harmonic on one of the factors which we denote by $M_j$.\footnote{For example, the Lichnerowicz-Obata theorem implies that exciting harmonics on two factors already produce a positive $\Delta_2$ eigenvalue, see Appendix \ref{app:delta2spectrum}.} This profile will have a negative $\Delta_2$ eigenvalue only if scalar nontrivial eigenfunctions on $M_j$ exist that satisfy
\beq
\Delta_0 < \frac{4\Lambda}{D-2}=\frac{2R_j}{D_j},
\eeq
where we used Einstein's equation to rewrite the condition in terms of $M_j$ data. For each such mode, the trace condition will reduce the free parameters to $n-1$. Therefore, if we introduce 
\beq
N^0_{M_j} = \left\{\text{$\#$ of scalar modes on $M_j\neq S^{D_j}$ satisfying $\frac{R_j}{D_j-1}<\Delta_0<\frac{2R_j}{D_j}$.} \right\},
\eeq
the number of such negative modes will be $(n-1) \cdot N_{M_j}^0$. As explained in Appendix \ref{app:delta2spectrum}, all these modes can be made transverse (by including a $\chi^{(j,j)}$ mode in the language of the Appendix) unless CKV are present that only happens when $M_j=S^{D_j}$. Since we already considered the case with spheres we will not repeat the analysis here.

\smallskip

The final result of our analysis is that the phase of the gravitational path integral on $M$ is given by
\bea
N_2 &=&n-1 + \sum_j N_2^{(j)}+ (n-1) \cdot \sum_{\text{$j$ for $M_j \neq S^{D_j}$}}N^0_{M_j}   \nonumber\\
&&+ (n-2)\cdot  \sum_{\text{$i$ for $M_i=S^{D_i}$ and $D_i>2$}}(D_i+1) .
\ea
The first factor of $n-1$ comes from the modes in \eqref{eq:YASUDANM}. The sum over $N_2^{(j)}$ counts the total number of negative tensor modes on $M_j$. The third term on the first line comes from all $h_{\mu\nu}$ constructed out of $M_j$ scalar modes leading to negative eigenvalues of $\Delta_2$. In the second line, we collect the contribution from those factors $M_i=S^{D_i}$ that are spheres using the results in Section \ref{sec:PRODSPHERE}. 

\smallskip

If the manifolds $M_j$ are all rigid, we can set $N_2^{(j)}=0$ but we cannot tell a priori the value of $N_{M_j}^0$ (unless all the factors are spheres). A simple nontrivial example we can find is the projective space. The eigenmodes are the same spherical harmonics but the scalar sector has $\ell=1$ removed implying that $N_{\mathbb{RP}}^0=0$. We can also check this for other rigid spaces using \cite{Bonifacio:2020xoc}. For example, $\mathbb{CP}^n$ also has no scalar mode in the range $R/(D-1) < \Delta_0 < 2R/D$. For products involving these manifolds, we can immediately claim
\beq
Z_{M}=(-\i)^{(n-2)\big(1+\sum_{\text{$M_i=S^{D_i>2}$}} (D_i+1)\big)}\,\cdot \,|Z_{M}|,
\eeq
for $M=M_1 \times \ldots \times M_n$ for $M_j = S^{D_j}, \mathbb{RP}^{D_j}$ or $\mathbb{CP}^{D_j/2}$. The phase of other spaces requires explicit knowledge of the first few scalar harmonics. If the product involves only two manifolds $M_1\times M_2$ in this class, then the path integral becomes real and positive.

\subsubsection{Charged black holes: Examples from Kaluza-Klein theory}\label{sec:MagneticNariai}

Generalizing the techniques in this paper to charged black holes is a straightforward but technically challenging problem. The graviton couples to the photon in nontrivial ways that lead to a more complicated action. It is not easy to reduce the evaluation of the phase, or even the absolute value of the partition function, to the knowledge of the spectrum of the Laplacian on the space. In this section, we bypass this problem by studying charged Kaluza-Klein black holes. 

\smallskip

For concreteness, we consider a theory of pure gravity in 5d and interpret it as a 4d theory of pure gravity coupled to an electromagnetic field (and other KK modes). Let us begin with the Einstein space $S^2 \times S^3$ with metric
\beq
\d s^2 = {\sf R}_1^2\,\, \Big( \sin^2\rho\, \d \eta^2 + \d \rho^2 \Big) + {\sf R}_2^2 \,\,\d \Omega_3^2, 
\eeq
with ${\sf R}_1^2 = {\sf R}^2/4$ and ${\sf R}_2^2 ={\sf R}^2/2$, and ${\sf R}$ is the 5d dS radius. This geometry arises from the Nariai limit of black holes in 5d, with $\eta$ being time and $\rho$ the radial coordinates between the black hole and cosmological horizon. Let us now parametrize the 3-sphere that appears in the metric as 
\beq
\d \Omega_3^2 = \frac{1}{4} \Big( \d \theta^2 + \sin^2 \theta \, \d \phi^2 + (\d\psi + \cos \theta\, \d \phi)^2 \Big),
\eeq
where $0\leq \theta\leq \pi$, $\phi \sim \phi + 2\pi$ and $\psi \sim \psi + 4\pi$. This is the representation of $S^3$ as a Hopf fibration. The 5d metric is therefore locally $S^2 \times S^2 \times S^1$ and we can perform a KK reduction on the $S^1$ parametrized by $\psi$. We end up with a 4d solution 
\beq
\d s^2 = \frac{{\sf R}_{\text{4d}}^2}{2}\,\, \Big( \sin^2\rho\, \d \eta^2 + \d \rho^2 \Big) + \frac{{\sf R}_{\text{4d}}^2}{4} \,\,\Big( \d\theta^2 + \sin^2\theta\,\d\phi^2\Big),~~~~~A=\cos\theta \, \d\phi , 
\eeq
where $A$ is the $\U(1)$ KK gauge field. The 4d dS radius is ${\sf R}_{\text{4d}} = {\sf R}/\sqrt{2}$, while the 4d Newton constant is $G^{\text{4d}}_N =G^{\text{5d}}_N/(2\pi {\sf R}_2)$. The geometry is therefore $\text{dS}_2 \times S^2$ with a magnetic field of unit charge. This is exactly what we were aiming for, although the size of the $S^1$ is comparable to the size of the base $S^2 \times S^2$ geometry.

\smallskip

Evaluating the path integral from the 4d geometry is complicated but now we can lift the solution back to 5d and apply our previous results. We already proved that the path integral on $S^2 \times S^3$ has a single unstable tensor mode. This fact, combined with the on-shell action, leads to
\beq
Z_M =\,\, e^{{\sf S}^{\text{4d}}_{\text{dS}}/2}\,\,\cdot\,\, |Z_M^{\text{quant}}|,
\eeq
a partition function that is real and positive. We wrote the on-shell action in terms of the 4d dS entropy ${\sf S}^{\text{4d}}_{\text{dS}}=\sqrt{2}\,\,{\sf S}^{\text{5d}}_{\text{dS}}$. Therefore, the partition function of a magnetically charged black hole in 4d is also real and positive. The action is also slightly subleading by a factor of $3/4$ compared to the uncharged Nariai contribution, and by a factor of $1/2$ compared to the dS$_4$ path integral. A thorough analysis of the physics of charged black holes in the Nariai limit can be found in \cite{Bousso:1996pn}\footnote{The unit magnetic charge black hole considered here has $|g|=1/3$ in the notation of that paper.} and more recently \cite{Castro:2022cuo}. Evaluating the on-shell action for general charges we can see that increasing the magnetic charge will only make the solution further subleading.

\subsubsection{Bound on the Lichnerowicz spectrum and application to $T^{pq}$}
The spectrum of the Lichnerowicz operator was extensively studied in the early days of AdS/CFT when considering AdS compactifications of string theory on Einstein spaces $AdS \times M$. These compactifications would produce a family of Kaluza-Klein modes with a mass dictated by the spectrum of $\Delta_2$ on $M$. In this context, it was important to verify whether modes that violate the BF bound on AdS were present. In this section we want to take some results from those studies and apply them to the case where $M$ is a Euclidean spacetime which solves Einstein's equation with a cosmological constant. In particular, one can derive a bound on the eigenvalues of $\Delta_2$ following \cite{Gibbons:2002th} and earlier work \cite{Page:1984ad,Gubser:2001zr}. Integrating by parts and using the commutation relations between covariant derivatives one can show, in any dimensions, that\footnote{A previous version had a typo in the index placement for the term involving the Riemann tensor, we thank Mukund Rangamani for pointing this out.}
\bea
\int_{M^D} h^{\mu\nu} \Delta_2 h_{\mu\nu}& =& \int_{M^D} \left[ 
 3(\nabla_{(\mu} h_{\nu\rho)})^2 - 4 R_{\mu\rho\sigma\nu}h^{\mu\nu}h^{\rho\sigma} + \frac{4 \Lambda}{D-2} h_{\mu\nu}h^{\mu\nu}\right],\nonumber\\
&\geq& \int_{M^D}\left[ - 4 R_{\mu\rho\sigma\nu} h^{\mu\nu} h^{\rho \sigma} + \frac{4 \Lambda}{D-2} h^{\mu\nu} h_{\mu\nu}\right]
\ea
A bound on the spectrum of $\Delta_2$ can be achieved by studying the spectrum of eigenvalues of the Riemann tensor defined as
\beq
R_{\mu\rho\sigma\nu} h^{\rho\sigma} = k h_{\mu\nu}.
\eeq
We can derive a bound now by choosing the maximum value of $k$, which we call $k_{\text{max}}$, and then
\beq
\Delta_2 \geq \frac{4 \Lambda}{D-2}- 4 k_{\text{max}}
\label{eq:tensorbound}
\eeq
We remind the reader that here $\Delta_2 = \Delta_L - 4\Lambda/(D-2)$ is the differential operator that appears in the quadratic Einstein action, which is shifted from what some references would refer to as the Lichnerowicz operator.

\subsubsection*{The path integral on $T^{pq}$}

In Section \ref{sec:MagneticNariai} we reinterpreted the gravitational path integral on $S^2 \times S^3$ as that of a magnetic black hole in $S^2 \times S^2$. This is a special case of a larger set of 5d Einstein spaces called $T^{pq}$ with $p,q$ co-prime integers. These are coset spaces $(\SU(2) \times \SU(2)) /\U(1)$. If the two $\SU(2)$ factors are generated by two sets of generators $\sigma_i$ and $\widetilde{\sigma}_i$ with $i=1,2,3$, the $\U(1)$ generator in the denominator is $p \sigma_3 + q \widetilde{\sigma}_3$. The metric on these spaces can be written explicitly \cite{Gubser:2001zr} and is given by
\bea
\d s^2 &=& {\sf R}_1^2\,\Big(\d \rho^2 + \sin^2\rho\,\d \eta^2\Big) + {\sf R}_2^2\,\Big(\d \theta^2 + \sin^2 \theta \d\phi^2\Big)\nonumber\\
&&+\,\,{\sf R}_3^2\,\Big(\d \psi - p \cos \rho \, \d \eta - q \cos \theta \, \d \phi\Big)^2.
\ea
We can find the three radii by solving Einstein's equation, which takes the form
\begin{equation}
\frac{2\Lambda}{3} = \frac{2{\sf R}_1^2 - p^2 {\sf R}_3^2}{2{\sf R}_1^4} = \frac{2{\sf R}_2^2 - q^2 {\sf R}_3^2}{2{\sf R}_2^4} = \frac{(q^2\,{\sf R}_1^4  + p^2\,{\sf R}_2^4 ) {\sf R}_3^2}{2{\sf R}_1^4 {\sf R}_2^4}.
\label{eq:placeholder}
\end{equation}
Upon dimensionally reducing on the circle $\psi$, this solution can be interpreted as the Euclidean section of a dyonic $\text{dS}_4$ black hole in the Nariai limit with magnetic charge $q$. To have a real electric charge would require an imaginary $p$ and we have not considered the problem of evaluating the phase of the path integral on a complex geometry (see \cite{Turiaci:2025xwi} for some comments). For this reason, we will restrict ourselves to thinking of the problem from the 5d point of view, instead of attempting to give a physical 4d interpretation.

\smallskip

The spectrum of scalar harmonics on $T^{pq}$ has been studied in the literature \cite{Gubser:1998vd,Ceresole:1999rq} but not the tensorial generalization, and the spectrum of $\Delta_2$ is not known. Although tedious, this is not a difficult calculation. In particular, the spectrum of harmonics on any coset space such as $T^{pq}$ can be obtained from group-theoretic considerations \cite{Pilch:1984xx}. Instead, similarly to \cite{Gubser:2001zr}, we will use it as an illustration to apply the bound \eqref{eq:tensorbound}.

\smallskip

Following the general strategy in section \ref{sec:strategy}, the scalar problem is trivial since we know that $T^{pq}$ has no CKV and therefore this sector only contributes a factor of $(+\i)$. We will use the bound derived earlier to explore whether $\Delta_2$ can have unstable modes. According to \eqref{eq:tensorbound}, the sufficient condition for the nonexistence of negative modes is 
\ie
k_{\text{max}}<\frac{1}{3}\Lambda
\label{eq:kbound},
\fe
where $k_{\text{max}}$ is the largest eigenvalue of the Riemann tensor. Importing the results of \cite{Gubser:2001zr} to our problem, we find:
\begin{itemize}
\item There is one mode that always exceed \eqref{eq:kbound}:
\ie
h_{ab}=\alpha (E_1^2+E_2^2)+\beta(E_3^2+E_4^2)+\gamma E_5^2,~~~~~\text{with}~~2\alpha+2\beta+\gamma=0
\\
k=\frac{1}{6} \left(\sqrt{49 \Lambda ^2-90 \Lambda +45}+\Lambda \right)>\frac{1}{3}\Lambda
\label{eq:negtensormode}
\fe
where
\begin{equation}
\begin{aligned}
& E^1={\sf R}_1 \d \rho, \quad E^2={\sf R}_1 \sin \rho \d\eta, \quad E^3={\sf R}_2 \d \theta, \quad E^4={\sf R}_2 \sin \theta \d\phi \\
& E^5={\sf R}_3\left(\d\psi-p \cos \rho \d\eta -q \cos\theta\d\phi\right)
\end{aligned}
\end{equation}
and
\ie
\alpha=&\,-\frac{8}{9} (4 \Lambda -3) \left(\sqrt{49 \Lambda ^2-90 \Lambda +45}+7 \Lambda -6\right)
\\
\beta=&\,\frac{4}{9} (4 \Lambda -3) \left(\sqrt{49 \Lambda^2-90 \Lambda +45}+7 \Lambda -3\right)
\\
\gamma=&\,\frac{4}{9} \left(\sqrt{49 \Lambda ^2-90 \Lambda +45}+\Lambda -3\right) \left(\sqrt{49 \Lambda^2-90 \Lambda +45}+7 \Lambda -3\right)
\fe
Its physical meaning is the relative change in size of the two $S^2$, with non-trivial fibering on $S^1$. Geometrically this is very similar to the negative mode of $S^2 \times S^2$. 
\item Another two modes can exceed \eqref{eq:kbound} for some $\Lambda$
\ie
h_{ab}=E_1 E_3-E_2E_4 ~~~\text{and} ~~~ h_{ab}=E_1 E_4-E_2E_3
\\
k=\frac{1}{2} \sqrt{-8 \Lambda ^2+18 \Lambda -9}>\frac{1}{3}\Lambda~~~~\text{when}~~0.801<\Lambda<1.33
\fe

\item All the other $15-3=12$ modes satisfy \eqref{eq:kbound}.
\end{itemize}
We find \eqref{eq:kbound} is violated. Is that sufficient to say that there exist negative modes? Actually, one can further show that the tensor in \eqref{eq:negtensormode} satisfies
\ie
\Delta_2 h_{ab}=\frac{2}{3}\left(\Lambda-\sqrt{49\Lambda^2-90\Lambda+45}\right)
\fe
One can show that the eigenvalue is negative in the full range of $\frac{1}{3}<\Lambda<\frac{2}{3}$.
Based on other examples we worked out, it is reasonable to guess that this is the single negative mode and the path integral on $T^{pq}$ is real and positive, but this requires further work to confirm.

\smallskip

We can apply the same bound to other Einstein manifolds. For example, the bound on $\Delta_2$ was applied in \cite{Gibbons:2002pq}, in the context of compactifications of supergravity, to the $D=7$ Einstein space $M^{pqr}$ which is a $\U(1)$ bundle over $\mathbb{CP}^2\times S^2$. While that reference found that for some choices of $p/q$ the compactification on $M^{pqr}$ can be stable, we can check that the same bound implies that the gravitational path integral on $M^{pqr}$ has $N_2\geq 1$ for all $p/q$. Another example is $Q^{n_1n_2n_3}$ which is a $\U(1)$ bundle over $S^2 \times S^2 \times S^2$ and we find that $N_2\geq 1$ for all $n$'s. This illustrates that the two notions of stability are physically different.

\subsubsection{Rotating black holes and the Page metric}

In this section, we analyze some aspects of the Kerr black hole in dS. We  consider possible smooth geometries that one can construct out of the Kerr metric, which might contribute to the same gravitational path integral as the sphere or the Nariai geometry. The explicit form of the metric for arbitrary mass $m$ and rotation $a$ parameters is
\begin{equation}
    \d s^2 = -\frac{\Delta_r}{\Xi^2\rho^2} \left[ \d t - a \sin^2 \theta \d\phi \right]^2 
    + \frac{\rho^2}{\Delta_r} \d r^2 + \frac{\rho^2}{\Delta_\theta} \d\theta^2 
    + \frac{\Delta_\theta \sin^2 \theta}{\Xi^2\rho^2} \left[ a \, \d t - (r^2 + a^2) \d\phi \right]^2 ,
\end{equation}
where we define $\Xi=1+a^2/{\sf R}^2$, $\rho^2 = r^2 + a^2 \cos^2 \theta$ and also the two functions
\beq
\Delta_r = (r^2 + a^2)\Big(1-\frac{r^2}{{\sf R}^2}\Big) -  \mu r,~~~~~~~~\Delta_\theta = 1+\frac{a^2}{{\sf R}^2} \cos^2 \theta,
\eeq
 and $0\leq \theta \leq \pi$ while $\phi \sim \phi+2\pi$. The horizons are located at the roots of $\Delta_r$. The largest root $r_c$ is the cosmological horizon. The next largest roots $0<r_-<r_+$ are the inner and outer horizons of the black hole. The cosmological and outer black hole horizons are not in thermal equilibrium in general. The two temperatures and angular velocities are 
\beq
\beta_{i} = \frac{4\pi \Xi (r_i^2+a^2)}{r_i\Big(1-\frac{3 r_i^2}{{\sf R}^2}-\frac{a^2}{{\sf R}^2}-\frac{a^2}{r_i^2}\Big)},~~~~\Omega_i = \frac{a}{r_i^2+a^2},
\eeq
with $i=+$ (black hole horizon) or $c$ (cosmological horizon). One can consider regimes of this metric where the outer black hole horizon approaches either its inner horizon (the extremal Kerr limit) or the cosmological horizon (the rotating Nariai limit) see \cite{Booth:1998gf, Anninos:2009yc}. Within this range, the geometry is not necessarily smooth when we consider its Euclidean continuation. In the rotating Nariai regime, there is a conical singularity at either horizon that can only be removed when $\beta_+ = \beta_c=\beta$. As discussed in \cite{Page:1978vqj}, having the temperatures match is not enough for smoothness since one also needs to impose 
\beq\label{eq:smoothOmega}
\beta_+\Omega_+ - \beta_c \Omega_c = \beta \left( \frac{a}{r_+^2+a^2} - \frac{a}{r_c^2+a^2}\right) = 2\pi \i n,
\eeq
with $n$ an integer (if the theory has no fermions) or an even integer (if the theory has fermions). One way to understand this condition is to consider two small loops, one around the black hole horizon and the other one with opposite orientation around the cosmological horizon. This is a contractible loop in the Euclidean geometry and the combined rotation measured by the angular velocities should be trivial. Only $n=0$ leads to a solution with a real Lorentzian continuation, but this implies that $r_c=r_+$ which is the Nariai limit. A more careful analysis leads to the conclusion that the only solution with $n=0$ also has $a=0$ and therefore reduces to the $S^2 \times S^2$ geometry arising from the Nariai limit of the non-rotating dS black hole.

\smallskip

Page found that the only other choice leading to a smooth geometry is $n=1$,\footnote{Notice that the fact that $\Delta(\beta \Omega) = 2\pi \i$ implies that the geometry has no spin structure. This means that if we make fermions be antiperiodic with respect to one horizon, say the cosmological one, they will unavoidably be periodic with respect to the other one, in this case the black hole horizon. A smooth spin structure requires $\Delta(\beta \Omega) = 4\pi \i$ but Page found this case to be singular. A possible way out is to consider charged black holes with appropriately charged fermions.  } while any other $n \geq 2$ leads to a singularity at the poles of the sphere parametrized by $\theta$ and $\phi$ \cite{Page:1978vqj}. Restricting to $n=1$ he applied a similar change of coordinates as the non-rotating case
\beq
r = r_0 - \epsilon \cos \rho ,~~~\eta = \frac{2\pi \tau}{\beta},~~~\phi \to \phi + \Omega_+ \tau,
\eeq
where $r_0= (r_++r_c)/2$ and $\epsilon=(r_c-r_+)/2$. In the $\epsilon \to 0$ limit we have $r_+ = r_c = r_0$. It is useful to study the geometry in terms of $r_0$ and $\nu=\i a/r_0$. In these variables the Nariai radius is $r_0 = \sqrt{(1+\nu^2)/(3-\nu^2)} {\sf R}$. The rescaled time has now period $\eta \sim \eta + 2\pi$. With the shift on $\phi$ implemented above, the contractible circle leaves $\phi$ unchanged. In the limit $\epsilon\to0$, the Kerr dS metric becomes
\bea
    \d s^2 &=& {\sf R}^2 (1 + \nu^2) \left[ \frac{1 - \nu^2 \cos^2 \theta}{3 + 6 \nu^2 - \nu^4} \, (\d \rho^2 + \sin^2 \rho \, \d \eta^2) \right. + \frac{1 - \nu^2 \cos^2 \theta}{(3 - \nu^2) - \nu^2 (1 + \nu^2) \cos^2 \theta} \, \d \theta^2 
    \nonumber\\
    &&+ \frac{(3 - \nu^2) - \nu^2 (1 + \nu^2) \cos^2 \theta}{(3 + \nu^2)^2 (1 - \nu^2 \cos^2 \theta)} \sin^2 \theta 
    \left. \left(\d \phi - \sin^2 \frac{\chi}{2}  \, \d\eta\right)^2 \right],
\ea
where $\nu = \i a/r_0$ is a dimensionless parameter. Imposing \eqref{eq:smoothOmega} fixes
\beq
\frac{4\nu(3+\nu^2)}{3+6\nu^2-\nu^4}=1,~~~\Rightarrow~~~\nu \approx 0.281702. 
\eeq
The sphere parametrized by $\rho$ and $\eta$ becomes dS$_2$ in Lorentzian signature, although the geometry is complex (as seen in the last term appearing in the metric). The geometry is therefore a nontrivial $S^2$ fibre bundle over a base $S^2$. The space can also be described as $\mathbb{CP}^2 \# \mathbb{CP}^2$. A generalization to higher dimensions can be found in \cite{Gibbons:2004uw}.

\smallskip

What is the phase of the gravitational path integral done over this geometry? The contribution of the scalar sector is $(+\i)$. In recent work \cite{Hennigar:2024gbg}, the authors analyzed the lowest lying states in the spectrum of $\Delta_2$ precisely for this metric. In order to do this they  focus first on metric perturbations that are diagonal and solely depend on $\theta$. By a numerical analysis of the problem, they found the first two lowest lying modes have eigenvalues which we call $\lambda_1<\lambda_2$ and are given by
\beq
\lambda_1 \approx - \frac{5.97}{{\sf R}^2} ,~~~~\lambda_2 \approx  \frac{11.79}{{\sf R}^2}. 
\eeq
Therefore in this sector there is a single negative mode. The authors of \cite{Hennigar:2024gbg} did not explore other metric fluctuation sectors. Nevertheless, we know from the general discussion in Appendix \ref{app:delta2spectrum} that this is the only relevant sector in geometries that take product forms. 

\smallskip 

For the reasons outlined above, it is reasonable to conjecture that the Page geometry has a single negative mode. If this is correct, the partition function on the Page geometry would be real and positive
\beq
Z_{\text{Page}}= \exp{\left( \frac{2(3-\nu^2)(1+\nu^2)^2}{(3+\nu^2)(3+6\nu^2-\nu^4)} \,{\sf S}_{\text{dS}}\right)}\, \cdot \, \big| Z^{\text{quant.}}_{\text{Page}}\big|
\eeq
Notice that the numerical coefficient appearing in the action is approximately $\approx 0.637\cdot {\sf S}_{\text{dS}}$ which is close but smaller than the $S^2 \times S^2$ contribution with an action $\approx 0.667\cdot {\sf S}_{\text{dS}}$. It would be interesting to show either analytically or numerically that there is a single unstable tensor mode in the geometry, as well as extending this analysis to the higher dimensional analog of the Page metric \cite{Gibbons:2004uw}.

\section{Discussion and future directions}\label{sec:Discussions}

In this paper, we have extended the formalism of \cite{Polchinski:1988ua,Maldacena:2024spf} and evaluated the gravitational path integral in a large class of spacetimes that solve Einstein's equations with a positive cosmological constant. We used some results in Riemannian geometry to reduce the problem to that of identifying physical instabilities of the geometry in the TT sector of metric fluctuations. We applied it to the Nariai black hole, generalizations with magnetic charge and angular momentum such as the Page metric, as well as other Einstein manifolds. We conclude with some discussion on future directions and conceptual issues raised by our calculations. 

\subsection{Comment on Maldacena's proposal}

Reference \cite{Maldacena:2024spf} analyzes the possibility that the phase of the path integral on the sphere might be canceled by incorporating an observer in dS. This resonates with some recent ideas that appeared in the context of the algebra of observables in dS \cite{Chandrasekaran:2022cip}. 

\smallskip

One of the main results in this paper is that the gravitational path integral on Einstein manifolds arising from (neutral charged or rotating) black holes in dS is real and positive. To some extent, this is an expected result. Adding a black hole in dS should be similar to adding an observer within pure gravity and therefore should be well-defined by itself. Nevertheless, the class of black holes that we considered is very special. The size of the black hole and cosmological horizons are comparable, and the horizon temperature is zero, i.e. the Nariai limit. So it is reasonable to ask what the consequences would be of explicitly incorporating an observer in this setup. 

\smallskip

Consider the motion of a particle on the manifold $S^2 \times S^{D-2}$. We parametrize the manifold as 
\ie
\d s^2=\sum_{i=1,2} {\sf R}_i^2\,\,\d\Omega^2_{D_i}=\sum_{i=1,2} {\sf R}_i^2\,\,(\cos^2\theta_i\,\d\tau_i^2+\d\theta_i^2+\sin^2\theta_i\,\d\Omega^2_{D_i-2})
\fe
For simplicity, we consider a very massive scalar—though not massive enough to form a black hole, so that its action can be evaluated using its proper length. A typical observer in the static patch of   $\text{dS}_{2}\times S^{D_2}$ (with $\tau_1\to \i t_1, \theta_1\to\arcsin r$) sits on a timelike curve at $r=0$ and fixed $\Omega_{D-2}$. The corresponding geodesic in $S^{2}\times S^{D-2}$ is a great circle on the first sphere while remaining at a fixed point on the remaining sphere: $\tau_1=s, \,\theta_1=0$ and $\Omega_{D-2}=0$. Expanding around the classical solution, we find
\ie
I=2\pi r_1 m+\frac{m r_1}{2}\int \d\tau_1\left[((\partial_{\tau_1}\theta_1)^2-\theta_1^2)+\frac{r_2^2}{r_1^2} (\partial_{\tau_1}\Vec{\Omega}_2)^2+\dots\right].
\fe
We find there is one negative mode corresponding to constant shifts of $\theta_1$, and also zero modes in constant shifts of $\vec{\Omega}_2$. If we generalize this to the geometry $\text{dS}_{D_1} \times S^{D_2} \times \ldots \times S^{D_n}$ there would be $D_1-1$ negative modes from the first factor and more copies of the zero-modes arising from the other spheres. The negative modes should contribute a factor of $(-\i)^{D_1-1}=(-\i)$. The final answer involves one extra factor of $+\i$ from the path integral of the clock mode \cite{Maldacena:2024spf}. The complete contribution from the observer is then
\beq
Z_{\text{observer}} = 1 \, \cdot \, |Z_{\text{observer}}|,
\eeq
where again we are only keeping track of the phase.

\smallskip

This result would suggest that the presence of an ``extra'' observer besides the black hole itself does not affect the final answer for the phase of the path integral which remains real and positive. (This is true for $S^2 \times S^{D-2}$ but not in the more general case of $S^{D_1} \times \ldots \times S^{D_n}$ which has a dimension-dependent phase.) Whether this is satisfactory or not depends on the interpretation one wants to give to the path integral on $S^2 \times S^{D-2}$, as explained in the introduction. The interpretation as a decay rate of dS into a pair of black holes requires an imaginary result which does not seem to be favored by this calculation. This implies that classic results on black hole creation during inflation have to be revisited, partly for this reason but also due to quantum corrections in their evaporation \cite{Bousso:1997wi,WOP_STW}.

\smallskip

Finally, we should mention that the calculation of the phase presents a large number of zero modes given by the location of the trajectory on the transverse spatial spheres. The overall phase can be changed by considering observers with ``angular momentum'' along those spheres.\footnote{This possibility was independently considered by Shoaib Akhtar, Raghu Mahajan and Haifeng Tang and we thank them for discussions.} Some of these zero modes can then become negative modes. It clearly deserves further work to determine what the correct prescription is and what the ultimate interpretation of the phase of these path integrals is.

\subsection{Application to black holes}

We focused on spaces with a positive cosmological constant in this paper. What happens if we evaluate the phase of the gravitational path integral in AdS? In particular, we can consider black hole solutions in pure gravity in asymptotically AdS spaces.

\smallskip

This first observation is that AdS has a conformal boundary, and the analysis has to be adapted. It is reasonable to expect that if we restrict the metric fluctuations to be normalizable in AdS and decay sufficiently fast at its boundary, then the manipulations in this paper still hold. In particular, we expect the contribution of vector modes to the phase to still be trivial. The contributions from the scalars are also trivial since now 
\beq
\widetilde{\Delta}_0^\gamma = \Big(\frac{2 \gamma(D-1)-(D-2)}{2}\Big) \Delta_0  + \frac{2 \gamma D}{D-2} |\Lambda|,
\eeq
is a strictly positive operator if $\Lambda<0$. This implies that the Wick rotation of the path integral that makes the Gibbons-Hawking-Perry negative mode introduces no phase to the partition function. These considerations imply that 
\beq
Z_{\text{AAdS}} = (-\i)^{N_2}\,\cdot\, |Z_{\text{AAdS}}|,
\eeq
where $N_2$ is the number of unstable TT modes in the geometry. Contrary to the dS case, this result is perfectly reasonable; we obtain an imaginary answer only when the geometry has an unstable direction and an interpretation of vacuum decay works in bulk as well as in the boundary following \cite{Coleman:1985rnk}. A flat space version of this was originally put forth by Gross, Perry and Yaffe in \cite{Gross:1982cv}. The incorrect assumption in \cite{GINSPARG1983245} was that this still holds in dS. Instead, we found in this paper that the phase from the scalar sector cancels the phase from the physical TT negative modes.

\smallskip

This result might be satisfactory in pure gravity, but extending it to black holes and other solutions beyond pure gravity is more challenging and interesting. For example, identifying the negative modes in Einstein-Maxwell theory is not as straightforward as in pure gravity, and it is not immediately obvious that the phase will solely get contributions from the unstable TT modes. Studying this question in the context of AdS/CFT or flat space black holes is an interesting direction.

\subsection{Other generalizations}

A very interesting technical generalization of the results in this paper would be to incorporate fermions, particularly in the case of black holes in AdS or flat space.

\smallskip

Relevant progress in the last years has been to develop a method to compute the Witten index of a black hole using smooth geometries that contribute to the gravitational path integral \cite{Cabo-Bizet:2018ehj,Iliesiu:2021are,Boruch:2022tno,Chen:2024gmc,Cassani:2024kjn,Boruch:2025qdq} (and this can also be extended to non-supersymmetric theories \cite{Chen:2023mbc}). These developments raise the following question. What is the phase of the gravitational path integral that computes the Witten index of the black hole? This would require incorporating fermions and in particular gravitini. The partition function should be real\footnote{This is already nontrivial since most of these solutions are themselves complex! Some preliminary comments on how to deal with this situation are in \cite{Turiaci:2025xwi}.}, but the overall sign matters; it determines whether the BPS black holes are mostly fermionic or bosonic. While the one-loop corrections around these geometries have been studied recently\cite{H:2023qko,Anupam:2023yns}, the phase and in particular the sign has never been computed. This is important since the overall sign predicted from gravity is something that can be compared with the result obtained from the field theory or quantum mechanical formulation dual to the black hole. 

\paragraph{Acknowledgments} We thank Shoaib Akhtar, Victor Ivo, Hari Kunduri, Guanda Lin, Raghu Mahajan, Juan Maldacena, Haifeng Tang, and Chih-Hung Wu for useful discussions. We thank Chih-Hung Wu for comments on a draft. GJT and XS are supported by the University of Washington and the DOE award DE-SC0011637. 

\appendix

\section{Analysis of the ghost sector}\label{app:ghosts}

In this Appendix, we go over the quadratic action of the ghost action. This will not be relevant for the evaluation of the phase of the path integral but does affect the absolute value. The ghosts arising from gauge-fixing diffeomorphisms are two anticommuting vectors $b_\mu$ and $c_\mu$ with an action
\beq
I_{\text{ghosts}} = \int \d^D x \, \sqrt{g}\,\, b_\mu \,\frac{\delta G^\mu[h+\nabla_{(\mu}\xi_{\nu)}]}{\delta \xi^\nu}\Bigg|_{\xi=0} c^\nu,
\eeq
Expanding to quadratic order
\beq
\delta^2 I_{\text{ghosts}} = -\int \d^D x\, \sqrt{g} \, b^\mu \left( \nabla^2 g_{\mu\nu} +  R_{\mu\nu} + (1-\beta) \nabla_\mu \nabla_\nu\right) c^\nu ,
\eeq
where the dots denote higher-order terms. We can apply the same Hodge decomposition we used earlier to the ghost sector, namely
\beq
c_\mu = {\sf c}_\mu + \nabla_\mu c,~~~~\nabla^\mu {\sf c}_\mu = 0.
\eeq
\beq
b_\mu = {\sf b}_\mu + \nabla_\mu b,~~~~\nabla^\mu {\sf b}_\mu = 0.
\eeq
In terms of these variables the ghost action, after some rearranging, is
\beq
\delta^2 I_{\text{ghosts}}=\int \d^4x\,\sqrt{g}\,\left[- {\sf b}_\mu \Delta_1 {\sf c}^\mu - \frac{1}{2 \gamma} b\tilde{\Delta}_0^\gamma\Delta_0\,c   \right].
\eeq
While all modes ${\sf c}_\mu$ and ${\sf b}_\mu$ participate in the path integral, not all scalar components $c$ and $b$ do. To determine which modes to exclude we compute the norm
\beq
\int \d^D x\, \sqrt{g}\, b_\mu c^\mu = \int \d^D x\, \sqrt{g}\, {\sf b}_\mu {\sf c}^\mu + \int \d^D x\, \sqrt{g}\, b \Delta_0 c.
\eeq
Therefore the only modes we should exclude are constant modes in $c$. The ghost path integral, keeping in mind its origin as a Jacobian \cite{Polchinski:1988ua}, is
\beq
Z_{\text{ghost}} = \frac{|\text{det}'\Delta_1| \, | \text{det}'\widetilde{\Delta}_0^\gamma|}{V}.
\eeq
The prime denotes the removal of the zero modes of $\Delta_1$ in the first factor. These are parametrized by Killing vectors of $M$ and correspond to diffeomorphisms that are not fixed by the gauge condition. Their contribution is corrected by including the factor of $V$, the volume of the isometry group of $M$, in the denominator. The prime in the second term denotes removal of zero modes of $\Delta_0$, which do not participate in the path integral. The operator $\widetilde{\Delta}_0^\gamma$ has negative and positive modes. In general, it does not have zero modes except for special values of $\gamma$. If it does have zero modes, which would be physical, they can be lifted simply by changing the value of $\gamma$. It is clear that the phase of the path integral is gauge invariant as long as $(D-2)/2(D-1)<\gamma<\infty$. What happens at the limiting values?

\smallskip

 Let us consider $\gamma\to\infty$ first. In this limit the operator $\widetilde{\Delta}_0^\gamma$ becomes
\beq
\lim_{\gamma\to\infty} \frac{1}{\gamma} \widetilde{\Delta}_0^\gamma = \widetilde{\Delta}_0.
\eeq
Therefore, the kinetic kernel for $h$ is $\widetilde{\Delta}_0$ while the kinetic kernel for $\chi$ is proportional to $\Delta_0 (\widetilde{\Delta}_0)^2$. This does not cause any particular issue for all Einstein spaces except for the sphere. The Yano-Nagano theorem implies that  $\widetilde{\Delta}_0$ develops a set of $D+1$ zero-modes in $h$ that are physical when $M=S^D$. For any other manifold $\widetilde{\Delta}_0$ has no zero modes so there is no issue. The resolution for the sphere was already explained in \cite{Polchinski:1988ua}. The action for the ghost scalar sector becomes proportional to $b \Delta_0 \widetilde{\Delta}_0 c $ and therefore the ghosts develop a zero-mode as well. The space of gauge transformations not fixed by the gauge fixing are the whole set of Killing vectors and CKV spanning $\SO(D+1,1)$, which has an infinite volume. On the other hand the kernel of $\widetilde{\Delta}_0$ spans the space of non-isometric CKV, namely $\SO(D+1,1)/\SO(D+1)$ which also has an infinite volume and corresponds to the $h$ integration. The divergence between the two volumes cancels in the path integral and only the volume of $\SO(D+1)$ remains reproducing the answer at finite $\gamma$.

\smallskip 

Let us now consider the lower bound $\gamma_{\text{min}}=(D-2)/2(D-1)$. In this case $\widetilde{\Delta}_0^{\gamma_{\text{min}}} = - D \Lambda/(D-1)$. This means that all modes of $h$ and $\chi$ are negative modes of $\widetilde{\Delta}_0^{\gamma_{\text{min}}}$! Now all modes of $h$ will be positive modes and there is no reason to rotate the contour. Instead, all modes of $\chi$ are negative modes so we need to rotate $\chi \to \i \chi'$. This would not introduce a phase other than the fact that we need to exclude the one mode in the kernel of $\Delta_0 \widetilde{\Delta}_0$ (or $D+2$ for the sphere). Therefore we recover the same phase as that for generic $\gamma$. The one-loop determinant is finite since the contributions from the functional determinant of $\widetilde{\Delta}_0^{\gamma_{\text{min}}}$ cancel in the final answer. In principle we could keep lowering $\gamma$ although the analysis in Section \ref{sec:strategy} should be adapted to the case where $\chi$ is the naturally negative mode and not $h$.  

\smallskip

The final situation to consider when there are accidental $\widetilde{\Delta}_0^\gamma$ zero modes. We can use $S^2 \times S^{D-2}$ as an example. The only zero mode of $\Delta_0\widetilde{\Delta}_0$ is the $(\ell_1,\ell_2)=(0,0)$ harmonic. The harmonics $(1,0)$ happen to be zero modes of $\widetilde{\Delta}_0^\gamma$ for $\gamma=1$ and the $(0,1)$ ones are zero mode when $\gamma=(D/2-1)^2$. The $(2,0)$ harmonics are zero modes when $\gamma=3/7$, etc. Actually, this even happens with $S^D$, the $\ell=2$ harmonics are zero modes when $\gamma=(D+1)(D-2)/(D^2+D-2)$. The possible divergence arising from these zero modes cancels in the one-loop determinant since the contribution from $\widetilde{\Delta}_0^\gamma$ always cancels between the metric scalars and the ghosts in the absolute value. Nevertheless, the choice of phase becomes subtle since we do not know a priori whether we should Wick rotate these zero modes or not. This can only be resolved by perturbing away from these values of $\gamma$, and keeping track of whether the modes become negative or positive modes.

\section{Lichnerowicz spectrum on $M_1\times\dots\times M_n$}
\label{app:delta2spectrum}

In this Appendix we derive the TT spectrum of $\Delta_2$ on $M=M_1\times\dots\times M_n$, where the $M_j$ are Einstein spaces with $\text{dim}(M_j)=D_j$ and $\text{dim}(M)=D=\sum_j D_j$. The product $M$ is itself an Einstein manifold only if $R_j/D_j=R/D$ for all $M_j$. The analysis generalizes that of \cite{Yasuda:1984py} to arbitrary spaces, and corrects some mistakes.

\smallskip

Let us begin by setting some notation. We denote the coordinates of the full geometry $M$ by greek indices $\mu,\nu, \ldots$. For each factor $M_j$, we introduce labels $\mu_j,\nu_j,\ldots$ that only run over the coordinates of $M_j$. We can represent the metric tensors on $M$ in terms of $n$-by-$n$ blocks with one index in $M_i$ and the second index in $M_j$ for $i,j=1,\ldots,n$. For example, in this language, the background metric on $M$ will be block diagonal with each block being the metric on $M_j$. We can express the metric fluctuation tensor on the product space in terms of these blocks
$$\phi_{\mu\nu}=\begin{pmatrix}
H_{\mu_1\nu_1} & \cdots & H_{\mu_1\nu_n}\\
\vdots &  & \vdots\\
H_{\mu_n\nu_1} & \cdots & H_{\mu_n\nu_n}\\
\end{pmatrix}.$$
This decomposition is useful because the differential operator $\Delta_2$ acting on $\phi_{\mu\nu}$ does not introduce mixing among blocks, namely
\ie
\,&\Delta_2 \phi_{\mu\nu}
=&\begin{pmatrix}
\ddots &  &  & \\
 & (\Delta_2^{(i)}+\sum_{j\neq i}\Delta_0^{(j)})H_{\mu_i\nu_i} &  \\
 &&& \cdots&(-\square^{(j)}-\square^{(k)}+\sum_{l\neq j,k}\Delta_0^{(i)})
 H_{\mu_j\nu_k}& \cdots\\
 & & &\ddots\\
\end{pmatrix},\nonumber
\fe
where  $\Delta_{2,1,0}^{(i)}$ denotes the tensor, vector, and scalar operator defined in Section \ref{sec:QuadraticAction} acting only on $M_i$. The reason behind the decoupling of the blocks with respect to $\Delta_2$ is that, on a product space, the components of the Riemann tensor vanish unless the four indices are on the same $M_j$.

\smallskip

We can apply the Hodge decomposition, similar to what we did in Section \ref{sec:QuadraticAction}, to each of the blocks $H_{\mu_i \nu_j}$ for $i,j=1,\ldots,n$. The diagonal tensor $H_{\mu_i \nu_i}$ can be decomposed into irreducible  tensors, vectors, and scalars on the subspace $M_i$
\ie
H_{\mu_i\nu_i} = \phi^{(i,i)}_{\mu_i\nu_i}+ \nabla_{\mu_i} \eta^{(i,i)}_{\nu_i} &+ \nabla_{\nu_i} \eta^{(i,i)}_{\mu_i} +2\Big(\nabla_{\mu_i}\nabla_{\nu_i}- \frac{1}{D_i} g_{\mu_i\nu_i}\square \Big) \chi^{(i,i)} + \frac{1}{D_i} g_{\mu_i\nu_i} h^{(i,i)}
\\
\text{with}~~~~
&\nabla^{\mu_i}\phi^{(i,i)}_{\mu_i\nu_i}=0,~~{\phi^{(i,i)\mu_i}}_{\mu_i}=0,~~\nabla^{\mu_i}\eta^{(i,i)}_{\mu_i}=0
\fe
where $\nabla_{\mu_i}$ is the covariant derivative on $M_i$. The superscript is used to indicate the block to which the components belong. The decomposition of the off-diagonal components mixing $M_j$ and $M_k$ with $j\neq k$ is given by
\ie
H_{\mu_j\nu_k}=\phi^{(j,k)}_{\mu_j\nu_k}+ \nabla_{\mu_j} \sigma^{(j)}_{\nu_k} + \nabla_{\nu_k} \sigma^{(k)}_{\mu_j}+2\nabla_{\mu_j} \nabla_{\nu_k}\zeta^{(j,k)}
\\
\text{with}~~~~\nabla^{\mu_j}\phi^{(j,k)}_{\mu_j\nu_k}=0,~~\nabla^{\nu_k}\phi^{(j,k)}_{\mu_j\nu_k}=0,~~\nabla^{\nu_j}\sigma^{(k)}_{\nu_j}=0.
\fe
The $\sigma^{(k)}_{\mu_i}$ are $n-1$ vectors along $M_i$. According to $\Delta_2$ these sectors are independent, but the TT condition on the full $\phi_{\mu\nu}$ will couple them. The trace-free condition is the simplest one
\beq
\sum_i h^{(i,i)} = 0,
\eeq
which we can use to eliminate one of the $h^{(i,i)}$ components. The transversality condition is more complicated
\ie
\Delta_1^{(i)}\eta^{(i,i)}_{\nu_i}+2\nabla_{\nu_i}\tilde{\Delta}_0^{(i)}\chi^{(i,i)}-\frac{1}{D_i}\nabla_{\nu_i}h^{(i,i)}+\sum_{j\neq i}(-{\nabla^{(j)}}^2\sigma_{\nu_i}^{(j)}+2\nabla_{\nu_i}\Delta_0^{(j)}\zeta^{(j,i)})=0
\fe
The vector and scalar components on the LHS have to vanish separately,
\ie
\Delta_1^{(i)}\eta^{(i,i)}_{\nu_i}-\sum_{j\neq i}{\nabla^{(j)}}^2\sigma_{\nu_i}^{(j)}=0
\\
\nabla_{\nu_i}\left(2\tilde{\Delta}_0^{(i)}\chi^{(i,i)}-\frac{1}{D_i}h^{(i,i)}+\sum_{j\neq i}2\Delta_0^{(j)}\zeta^{(j,i)}\right)=0
\label{eq:transversality}
\fe
These equations can be used to eliminate $\eta_{\nu_i}^{(i,i)}$ and $\chi^{(i,i)}$. One can verify that the total number of remaining degrees of freedom is 
$\frac{1}{2}D(D-1)-1$, which matches the count for a TT tensor in $D$-dimensions. 

\smallskip

The original path integral is over $\phi_{\mu\nu}$ and some of the degrees of freedom introduced above are excluded. For example, if a scalar mode is independent of the coordinates on some $M_j$ then $\chi^{(j,j)}$ and all $\zeta^{(j,k)}$ with $k\neq j$ are excluded. This also implies that the scalar component of \eqref{eq:transversality} along $M_j$ is trivially satisfied since $\nabla_{\mu_j} h^{(j,j)}=0$ by assumption, removing one of the $n$ equations. If the scalar mode is a CKV along $M_j$, which can only happen when it is a sphere, then only $\chi^{(j,j)}$ is excluded, and the number of equations is not reduced. Finally, $\zeta$ modes that only depend on the coordinates of one of the factors $M_i$ are also excluded. In the vector sector, if $\eta_{\mu_i}^{(i,i)}$ is a Killing vector of $M_i$ then it is excluded but that same mode on $\sigma^{(k)}_{\mu_i}$ is not. $\sigma^{(k)}_{\mu_i}$ is only excluded if the mode is independent of $M_k$. Moreover, a mode that is a Killing vector along $M_i$ and depends on a single $M_k$ is also excluded since it cannot satisfy transversality.

\smallskip

We take a more careful look at the action of the differential operator $\Delta_2$ on these independent modes introduced above:
\begin{itemize}
\item For the irreducible tensor that appears in the diagonal block $\phi^{(i,i)}_{\mu_i\nu_i}$, it is easy to show that 
$$
\Delta_2 \phi^{(i,i)}_{\mu_i\nu_i} = \left( \Delta_2^{(i)} +\sum_{j\neq i}\Delta_0^{(j)} \right)\phi^{(i,i)}_{\mu_i\nu_i}.
$$
This structure indicates that the $\Delta_2$ eigentensor on the product space is given by the eigentensor of $\Delta_2^{(i)}$ on $M_i$, multiplied by scalar eigenmodes on the other submanifold. Since $\Delta_0^{(j)}$ is non-negative, $\phi^{(i,i)}_{\mu_i\nu_i}$ has a positive spectrum if $M_i$ is a rigid Einstein manifold.

\item A similar consideration applies to the diagonal mode $\eta^{(i,i)}_{\mu_i}$ since they satisfy
\beq\label{eq:DElTA2ONETA}
\Delta_2 \, \nabla_{\mu_i} \eta^{(i)}_{\nu_i} = \nabla_{\mu_i} \Big(\Delta_1^{(i)} + \sum_{j\neq i} \Delta_0^{(j)} \Big) \eta^{(i)}_{\nu_i},
\eeq
which can be derived by applying the Ricci identity to commute covariant derivatives. The same expression is valid for the $\sigma_{\mu_i}^{(k)}$ modes and therefore they have the same spectrum. The eigentensors will then be made of a product of a vector harmonic along $M_i$ multiplied by scalar harmonics in the other factors. We already showed in section \ref{sec:strategy} that the spectrum of $\Delta_1$ is positive so we do not need to worry about these modes. 

\item We look now at the scalar modes on the diagonal blocks. For the trace $h^{(i,i)}$, we have
\ie
\Delta_2 \, g_{\mu_i\nu_i}h^{(i,i)}=g_{\mu_i\nu_i}\Big(\sum_{j}\Delta_0^{(j)}-\frac{4\Lambda}{D-2}\Big)h^{(i,i)}
\label{eq:tracemode}
\fe
We can construct an $\Delta_2$ eigentensor by taking $h^{(i,i)}$ be a product of scalar eigenmodes in all $M_j$. The other scalar mode in the diagonal $\chi^{(i,i)}$ has the same spectrum. To show this we start from the identity 
\beq\label{eq:identity}
\nabla^2 \nabla_\mu \nabla_\nu f = \nabla_\mu \nabla_\nu \nabla^2 f + (R_{\nu \beta}g_{\mu \alpha}+R_{\mu\alpha}g_{\nu \beta}-2R_{\alpha \mu \beta \nu})\nabla^\alpha \nabla^\beta f
\eeq
This relation holds only for Einstein spaces, since otherwise there are extra terms proportional to $\nabla_\mu R_{\nu\rho}$. This identity implies that
$$
\Delta_2 \Big(\nabla_{\mu_i}\nabla_{\nu_i}- \frac{1}{D_i} g_{\mu_i\nu_i}\square \Big) \chi^{(i,i)} = \Big(\nabla_{\mu_i}\nabla_{\nu_i}- \frac{1}{D_i} g_{\mu_i\nu_i}\square \Big)\Big(\sum_{j}\Delta_0^{(j)}-\frac{4\Lambda}{D-2}\Big) \chi^{(i,i)}.
$$
For each scalar eigenmode in $h^{(i,i)}$ we can construct another metric fluctuation with the same eigenmode for $\chi^{(i,i)}$, with the same $\Delta_2$ eigenvalue, and use \eqref{eq:transversality} to form a linear combination that is TT. The subtleties that appear in the presence of CKV are considered in Section \ref{sec:PRODSPHERE}.

We can analyze which of these scalar modes are negative. The trivial constant mode is excluded from $\chi^{(i,i)}$ and $\zeta^{(i,j)}$ but included in $h^{(i,i)}$ and is obviously TT. The eigenvalue of $\Delta_2$ with $h^{(i,i)}$ constant is 
\ie
\,&\Delta_2 \begin{pmatrix}
\ddots &  &  &  0\\
 & g_{\mu_i\nu_i}c_i &  \\
 0& & &\ddots\\
\end{pmatrix} = - \frac{4\Lambda}{D-2} \, \begin{pmatrix}
\ddots &  &  &  0\\
 & g_{\mu_i\nu_i}c_i &  \\
 0& & &\ddots\\
\end{pmatrix},~~~\sum_i D_i c_i =0.
\fe
This leads to $n-1$ universal negative modes identified in \cite{Yasuda:1984py} appearing in all products of Einstein spaces. 

The next possibility is for $h^{(i,i)}$ to be a non-trivial mode on $M_j$ and be independent of other $M_k$ for all other $k\neq j$. This excludes all $\zeta$ and $\chi$ modes  except $\chi^{(j,j)}$ (again, the case with CKV is special and was analyzed in the main text). The Lichnerowicz-Obata theorem implies that they satisfy
\beq
\Delta_0^{(j)} - \frac{4\Lambda}{D-2} \geq - \frac{2(D_j-2)}{(D-2)(D_j-1)}.
\eeq
This mode can potentially be negative, unless $D_j=2$. To determine whether the first excited mode is negative or not requires precise knowledge of the scalar spectrum on $M_j$.  If none of the $M_i$ are spheres, there are therefore $(n-1) N_j$ negative modes where $N_j$ is the number of non-trivial scalar modes with $\Delta_0^{(j)} <4\Lambda/(D-2)=2 R_j/D_j$. The case where some of the $M_j$ are spheres was considered in Section \ref{sec:PRODSPHERE} and requires extra care due to the CKV.

The next possibility is for $h^{(i,i)}$ to be non-trivial along two spaces $M_i$ and $M_j$ and trivial along $M_k$ for all other $k\neq i,j$. In this case the Lichnerowicz-Obata theorem implies that 
\beq\label{eq:2modeLO}
\Delta_0^{(i)} + \Delta_0^{(j)} - \frac{4 \Lambda}{D-2} \geq \frac{2(D_i+D_j-2)}{(D-2)(D_i-1)(D_j-1)} >0.
\eeq
which is automatically positive. We do not need to consider the scalar sector further since all modes will be positive.  

\item The differential operator in the off-diagonal block is positive. This can be mostly easily seen from the form of $\Delta_2$. Since the off-diagonal Riemann tensor is zero, the Laplacian left in $\Delta_2$ is
\beq
\Delta_2 \, H_{\mu_i \nu_j} = - \nabla^2 H_{\mu_i \nu_j},~~~~\text{for $i\neq j$},
\eeq
and guarantees the non-negativity. Concretely, the off-diagonal transverse tensor $\phi^{(j,k)}_{\mu_j\nu_k}$ is given by the tensor product of the eigenvector in $M_j$ and $M_k$\footnote{We can verify this through the counting of degrees of freedom. The two transverse conditions remove $D_j$ and $D_k$ modes from the $D_j\times D_k$ matrix. These $D_j+D_k$ are not all independent since the double divergence  $\nabla^{\mu_j}\nabla^{\nu_k}\phi^{(j,k)}_{\mu_j\nu_k}=0$ gives a repeated constraint. This leaves $D_j\times D_k-D_j-D_k+1=(D_j-1)(D_k-1)$ degrees of freedom, which is the product of that of two transverse vectors.}, multiplied by scalar eigenfunctions on the other submanifolds.

Metric fluctuations constructed out of $\sigma_{\mu_i}^{(j)}$  satisfy \eqref{eq:DElTA2ONETA} and diagonalize $\Delta_2$ if they are an eigenvector on $M_i$ multiplied by scalar eigenfunctions on the rest of the submanifolds. They combine with $\eta_{\nu_i}^{(i,i)}$, which have the same form and $\Delta_2$-eigenvalue, to construct TT modes on $M$. If the eigenvector along $M_i$ is not a Killing vector, these modes can be parametrized by $n-1$ independent $\sigma^{(j)}_{\mu_i}$. If the eigenvector along $M_i$ is a  Killing vector  then the TT condition gives one constraint on $\sigma^{(j)}_{\mu_i}$ leaving $n-2$ independent modes.\footnote{This possibility was incorrectly excluded in \cite{Yasuda:1984py}.} This is true unless the mode is independent of some $M_k$ with $k\neq i$ since in that case $\sigma^{(k)}_{\mu_i}$ is excluded from the spectrum. The mode that is a Killing vector on $M_i$ and a constant on all other submanifolds is not part of the spectrum of metric fluctuations and should be excluded.

It is instructive to look at the off-diagonal scalar sector since one might think they potentially lead to negative modes. Using \eqref{eq:identity} we can derive
\bea
\Delta_2 \nabla_{\mu_j} \nabla_{\nu_k} \chi^{(j,k)} &=&\nabla_{\mu_j} \nabla_{\nu_k} \Delta_0 \chi^{(j,k)}  - (g_{\mu_j \alpha} R_{\nu_k \beta} + R_{\mu_j \alpha} g_{\nu_k\beta} ) \nabla^\alpha \nabla^\beta \chi^{(j,k)} ,\nonumber\\
&=&\nabla_{\mu_j} \nabla_{\nu_k} \left(\Delta_0-\frac{4\Lambda}{D-2}\right) \chi^{(j,k)}.
\ea
Therefore, the eigenvalue under $\Delta_2$ of a scalar eigenmode $\chi^{(j,k)}$ is the same as if that same mode would appear in $h^{(i,i)}$ or $\chi^{(i)}$.  How is this consistent with the observation that $\Delta_2$ acting on the off-diagonal block is a positive operator? The reason is that for a mode $\chi^{(j,k)}$ to participate in the gravitational path integral, it has to be non-trivial in at least two subspaces $M_j$ and $M_k$ with $k\neq j$. Otherwise, the combination $\nabla_{\mu_j} \nabla_{\nu_k} \chi^{(j,k)}$ would vanish. The Lichnerowicz-Obata theorem applied to $M_j$ and $M_k$ puts a lower bound \eqref{eq:2modeLO} which implies that there are no negative or zero modes in this sector. This is consistent with the earlier observation that no off-diagonal block can lead to negative modes.

\end{itemize}

To obtain the full one-loop answer, we need to evaluate the path integral measure for each mode descending from the ultralocal measure. The inner product defined in section \ref{sec:PIM} should be evaluated on all surviving modes, after we solve TT conditions. Nevertheless, we already know that the answer should reduce to the product of $\Delta_2$ eigenvalues, given the analysis in Section \ref{sec:PIM}. This is the final answer regardless of how we decide to parametrize the modes, but it might be useful to write the full one-loop determinant explicitly in terms of functional determinants on each individual $M_j$. This can be done with the information collected here but also requires computing the spectrum of $\Delta_1$ on $M$ and we leave it for future work.

\bibliographystyle{utphys2}
{\small \bibliography{Biblio}{}}

\end{document}